\documentclass[10pt]{article}

\usepackage[hyphens]{url}
\usepackage{graphicx}
\usepackage[utf8]{inputenc}
\usepackage{textcomp,marvosym}
\usepackage{amsmath,amssymb}
\usepackage{cite}
\usepackage{nameref,hyperref}
\usepackage{authblk}
\usepackage{multirow}

\begin{document}
\title{Computational Screening of Tip and Stalk Cell Behavior Proposes a Role for Apelin Signaling in Sprout Progression}

\author[1,a]{Margriet M. Palm}
\author[2]{Marchien G. Dallinga}
\author[1,b]{Erik van Dijk}
\author[2]{Ingeborg Klaassen}
\author[2]{Reinier O. Schlingemann}
\author[1,3]{Roeland M.H. Merks\thanks{merks@cwi.nl}}

\affil[1]{Life Sciences Group, Centrum Wiskunde \& Informatica, Amsterdam, the Netherlands}
\affil[2]{Ocular Angiogenesis Group, Academic Medical Center, Amsterdam, the Netherlands}
\affil[3]{Mathematical Institute, Leiden University, Leiden, the Netherlands}
\affil[a]{Present address: Division of Toxicology, LACDR, Leiden University, Leiden, The Netherlands}
\affil[b]{Present address: Center for Integrative Bioinformatics, VU University, Amsterdam, the Netherlands}

\date{\today}

\maketitle

\section*{Abstract}
Angiogenesis involves the formation of new blood vessels by sprouting or splitting of existing blood vessels. During sprouting, a highly motile type of endothelial cell, called the tip cell, migrates from the blood vessels followed by stalk cells, an endothelial cell type that forms the body of the sprout. To get more insight into how tip cells contribute to angiogenesis, we extended an existing computational model of vascular network formation based on the cellular Potts model with tip and stalk differentiation, without making a priori assumptions about the differences between tip cells and stalk cells. To predict potential differences, we looked for parameter values that make tip cells (a) move to the sprout tip, and (b) change the morphology of the angiogenic networks. The screening predicted that if tip cells respond less effectively to an endothelial chemoattractant than stalk cells, they move to the tips of the sprouts, which impacts the morphology of the networks. A comparison of this model prediction with genes expressed differentially in tip and stalk cells revealed that the endothelial chemoattractant Apelin and its receptor APJ may match the model prediction.  To test the model prediction we inhibited Apelin signaling in our model and in an \emph{in vitro} model of angiogenic sprouting, and found that in both cases inhibition of Apelin or of its receptor APJ reduces sprouting.  Based on the prediction of the computational model, we propose that the differential expression of Apelin and APJ yields a ``self-generated'' gradient mechanisms that accelerates the extension of the sprout.

\section*{Introduction}

Angiogenesis, the formation of new blood vessels from existing vessels, is important in numerous mechanisms in health and disease, including wound healing and tumor development. As a natural response to hypoxia, normal cells and tumor cells secrete a range of growth factors, including vascular endothelial growth factors (VEGFs)  and fibroblast growth factors (FGFs). These activate quiescent endothelial cells to secrete proteolytic enzymes, to migrate from the blood vessel and organize into an angiogenic sprout. Angiogenic sprouts are led by tip cells, a highly migratory, polarized cell type that extends numerous filopodia \cite{Gerhardt2003a}. Tip cells express high levels of the VEGF receptor VEGFR2 \cite{Gerhardt2003a}, Delta-like ligand 4 (Dll4)  \cite{Claxton2004} and, \emph{in vitro}, CD34 \cite{Siemerink2012}. The tip cells are followed by stalk cells \cite{Gerhardt2003a}, a proliferative and less migratory type of endothelial cell, which expresses low levels of Dll4 \cite{Claxton2004} and, \emph{in vitro}, have undetectable levels of CD34 \cite{Siemerink2012}  

The behavior of tip and stalk cells during angiogenic sprouting has been well characterized in mouse retina models and in endothelial spheroids \cite{Jakobsson2010,Arima2011}. From a mechanistic point of view, however, it is not well understood why two types of endothelial cells are involved in angiogenesis. Experimental and computational lines of evidence suggest that in absence of tip and stalk cell differentiation, endothelial cells can form blood-vessel like structures, albeit with abnormal morphological parameters.  In cell cultures, endothelial cells organize into network-like structures, without obvious differentiation into tip and stalk cells \cite{Folkman:1980vf,Califano:2008ct}, although the individual endothelial cells were found to vary in other aspects of their behavior, e.g., their tendency to occupy the nodes of vascular networks \cite{Parsa:2011ge}. Computational models have suggested a range of biologically-plausible mechanisms, by which populations of identical endothelial cells can self-organize into vascular network-like structures \cite{Manoussaki1996,Serini2003,Merks2006b,Merks2008,Oers2014,Szabo2007,Szabo2008} and sprout-like structures can form from endothelial spheroids \cite{Merks2008,Szabo2010,Oers2014}. Experimental interference with tip and stalk cell differentiation modifies, but does not stop the endothelial cells' ability to form networks.  In mouse retinal vascular networks, inhibition of Notch signaling increases the number of tip cells and produces denser and more branched vascular networks \cite{Hellstrom2007,Suchting2007a,Lobov2007a}, while in gain-of-function experiments of Notch the fraction of stalk cells is increased, producing less extensive branching \cite{Hellstrom2007}. \emph{In vitro}, similar effects of altered Notch signaling are observed \cite{Sainson2005,Williams2006a,Scehnet2007}. Taken together, these observations suggest that differentiation between tip and stalk cells is not required for vascular network formation or angiogenic sprouting. Instead they may fine-tune angiogenesis, e.g., by regulating the number of branch points in vascular networks. 

The exact mechanisms that regulate the differentiation of tip and stalk cell fate are subject to debate. Activation of the VEGFR2 by VEGF-A, which is secreted by hypoxic tissue, upregulates Dll4 expression \cite{Lobov2007a,Ridgway2006a,Patel2005,Hainaud2006}. Dll4 binds to its receptor Notch in adjacent endothelial cells, where it induces the stalk cell phenotype \cite{Jakobsson2009}, which includes downregulation of Dll4. The resulting lateral inhibition mechanism, together with increased VEGF signaling close to the sprout tip, may stimulate endothelial cells located at the sprout tip to differentiate into tip cells ``in place''. Detailed fluorescent microscopy of growing sprouts \emph{in vitro} and \emph{in vivo} shows that endothelial cells move along the sprout and ``compete'' with one another for the tip position \cite{Jakobsson2010,Arima2011}. Endothelial cells expressing a lower amount of VEGFR2, and therefore producing less Dll4, are less likely to take the leading tip cell position, while cells that express less VEGFR1, which is a decoy receptor for VEGFR2 \cite{Hiratsuka1998,Rahimi2000a}, are more likely to take the tip cell position \cite{Jakobsson2010}. These results suggest that the VEGF-Dll4-Notch signaling loop is constantly re-evaluated and thereby tip cell fate is continuously reassigned.  A series of recent observations, however, support an opposing view in which tip cells differentiate more stably.  Tip cells express the sialomucin CD34, making it possible to produce ``tip cell'' (CD34+) and ``stalk cell'' (CD34-) cultures using fluorescence-activated cell sorting (FACS) \cite{Siemerink2012}.  CD34+ cells have a significantly lower proliferation rate than CD34- cultures during the first 48 hours, suggesting that during this time they do not redifferentiate into stalk cells. In cultures of CD34-negative endothelial cells (stalk cells), the wild-type ratio of tip and stalk cells reestablishes only after around ten days. Thus within the time frame of \emph{in vitro} vascular network formation of around 24 to 48 hours \cite{Arnaoutova2009} cross-differentiation between tip and stalk cells is relatively rare. These data suggest that the differentiation between tip and stalk cells depends on a balance between (a) lateral inhibition via the Dll4-Notch pathway \cite{Hellstrom2007,Suchting2007a,Siekmann2007,Lobov2007a}, and (b) a stochastically ``temporary stabilized'' tip or stalk cell fate, potentially correlated with CD34 expression \cite{Siemerink2012}.

To develop new hypotheses on the role of tip and stalk cell differentiation during angiogenesis, we developed an explorative approach inspired by Long \emph{et al.} \cite{Long2012} who used a genetic algorithm to identify the transition rules between endothelial cell behaviors that could best reproduce \emph{in vitro} sprouting. Here we use a cell-based, computational model of angiogenesis \cite{Merks2008} that is based on the cellular Potts model (CPM) \cite{Graner1992,Glazier1993a}.  We extend the model with tip and stalk cell differentiation, and systematically vary the parameters of the tip cells to search for properties that make the ``tip cells'' behave in a biologically realistic manner: i.e., they should move to the sprout tip and affect the overall branching morphology.  We consider both a ``pre-determined'' model in which endothelial cells are stably differentiated into tip and stalk cells throughout the simulation time of the model,  and a ``lateral inhibition'' model, in which tip and stalk cells cross-differentiate rapidly via Dll4-Notch signaling.  We compare the  tip cell properties that our model predicts with differential gene expression data, and perform initial experimental tests for the resulting gene candidate \emph{in vitro}.

\section*{Results}

To develop new hypotheses on the role of tip cells during angiogenesis, we took the following ``agnostic'' approach that combines bottom-up modeling, bioinformatical analysis and experimental validation. We started from a previously published computational model of de novo vasculogenesis and sprouting angiogenesis \cite{Merks2008}. Briefly, the model simulates the formation of sprouts and vascular networks from a spheroid of identical ``endothelial cells'', driven by an autocrine, diffusive chemoattractant that drives endothelial cells together (see Ref.~\cite{Merks2008} and Materials and Methods for details). In the first step, we assumed that a fraction of the cells are ``tip cells'' (tip cell fraction) and the remaining cells are ``stalk cells'', hence assuming that cross-differentiation between tip and stalk cells does not occur over the course of the simulation. We next systematically varied the model parameters of the tip cells to look for cell behavior that (a) takes the tip cells to the sprout tips, and (b) changes the morphology of the simulated vascular networks formed in the model. The predicted differences between tip cell and stalk cell behavior were then expressed in gene ontology terms, so as to compare them with published gene expression differences between tip and stalk cells \cite{Siemerink2012}. The analysis yielded a gene candidate that was further tested in an \emph{in vitro} model of spheroid sprouting.  

As a computational model for angiogenesis, we used our previous cell-based model of de novo vasculogenesis and sprouting angiogenesis \cite{Merks2008}. The model assumes that endothelial cells secrete an autocrine, diffusive chemoattractant to attract one another. Due to the resulting attractive forces between the endothelial cells, the cells aggregate into a spheroid-like configuration. If the chemoattractant sensitivity of the endothelial cells is restricted to the interfaces between the endothelial cells and the surrounding ECM by means of a contact inhibition mechanism, the spheroids sprout in microvascular-network-like configurations. Although our group \cite{Merks2006b,Palm2013,Oers2014} and others \cite{Manoussaki1996,Tranqui2000,Gamba2003,Serini2003,Namy2004,Szabo2007} have suggested numerous plausible alternative mechanisms for de novo vasculogenesis and sprouting, in absence of a definitive explanatory model of angiogenesis we have selected the contact inhibition model for pragmatic reasons: It agrees reasonably well with experimental observation \cite{Merks2008,Gory-Faure1999}, it focuses on a chemotaxis mechanism amenable to genetic analysis, and it has a proven applicability in studies of tumor angiogenesis \cite{Shirinifard2009}, age-related macular degeneration  \cite{Shirinifard2012},  and toxicology \cite{Kleinstreuer2013}.

The computational model is based on a hybrid, cellular Potts and partial differential equation model \cite{Graner1992,Glazier1993a,Savill1996}. The cellular Potts model (CPM) represents biological cells as patches of connected lattice sites on a finite box $\Lambda$ of a regular 2D lattice $\Lambda\subset\mathbb{Z}^{2}$  with each lattice site $\vec{x}\in\Lambda$  containing a {\em cell identifier} $\sigma\in\mathbb{Z}^{+,0}$ that uniquely identifies each cell. Each cell $\sigma$ is also associated with a cell type $\tau(\sigma)\in\{\text{tip},\text{stalk},\text{ECM}\}$.  To mimic amoeboid cell motility the method iteratively attempts to move the interfaces between adjacent cells, depending on the amplitude of active membrane fluctuations (expressed as a ``cellular temperature'' \cite{Angelini2011} $\mu(\tau)$) and on a force balance of the active forces the cells exert on their environment (e.g. due to chemotaxis or random motility) and the reactive adhesive, cohesive and cellular compression forces. Assuming overdamped motility, the CPM solves this force balance as a Hamiltonian energy minimization problem (see Materials and Methods for details). 

The angiogenesis model includes the following endothelial cell properties and behaviors: cell-cell and cell-matrix adhesion, volume conservation, cell elasticity, and chemotaxis at cell-ECM interfaces. To describe cell-cell adhesion we define a contact energy $J(\tau,\tau^\prime)$  that represents the interfacial tension between cells of type $\tau$ and $\tau^\prime$. This term lumps contributions due to cell-cell adhesion \cite{Bentley2014} and cortical tensions \cite{Krieg:2008jd}. We assume that cells resist compression and expansion by defining a resting area $A(\tau)$. In practice the cells fluctuate slightly around their resting area depending on the elasticity parameter $\lambda(\tau)$. The cells secrete a diffusive chemoattractant $c$ at a rate $\alpha(\tau)$, with $\frac{\partial c}{\partial t}=D\nabla^2 t  - \epsilon(\tau) c + \alpha(\tau)$, where $D$ is a diffusion coefficient, $\epsilon$ is a degradation rate, which is zero inside cells, and $\alpha(\text{ECM})=0$. Chemotaxis at cell-ECM interfaces is incorporated by biasing active cell extension and retractions up chemoattractant with a factor $\chi(\tau)$, which is the chemoattractant sensitivity. 

\begin{figure}[!htbp]
  \centering
\includegraphics[width=\textwidth]{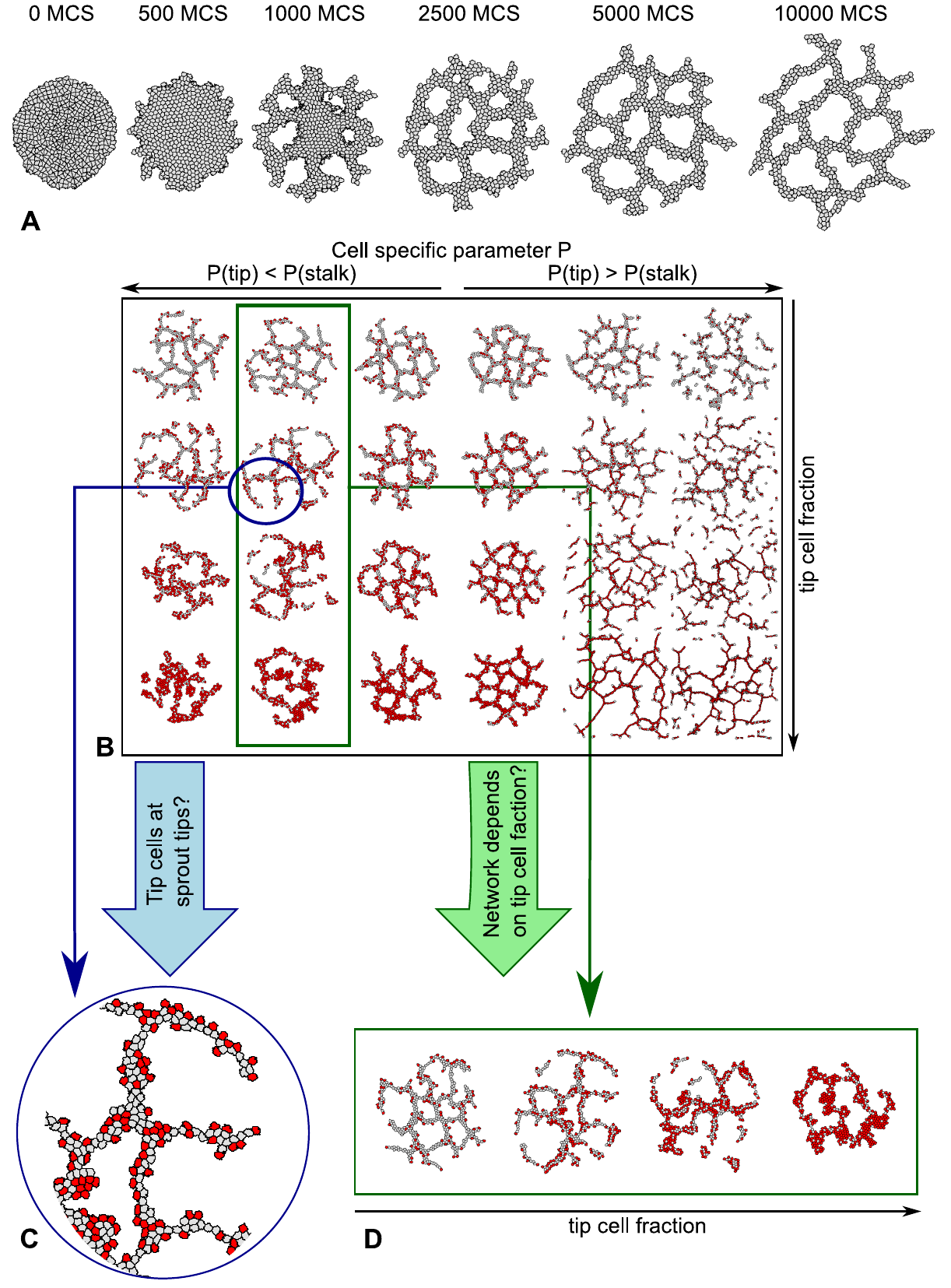}
\caption{\textbf{Overview of the angiogenesis model and the parameter search.} \textbf{A} Time-lapse of angiogenesis model behavior \textbf{B} For each parameter P that is tested in the parameter search a morphospace is created to compare the different parameter values for different tip cell fractions. \textbf{C} Each morphology is studied in detail to see if the sprout tips are occupied by tip cells (red). \textbf{D} Each row of morphologies is studied to find rows in which the morphologies differ, indicating that network formation depends on the tip cell fraction.}
  \label{fig:parsweep}
\end{figure}

We start the analysis from the set of nominal parameters listed in Table~\ref{tab:par}; these yield the nominal collective cell behavior shown in Fig ~\ref{fig:parsweep}\textbf{A}. The parameters are set according to experimental values as far as possible. The cross-sectional area of the endothelial cells in the cell cultures was $360 \pm 100\mu m^2$ (see Materials and Methods for detail), based on which we set the target area of the cells, $A(\text{tip})$ and $A(\text{stalk})$, to 100 lattice sites, corresponding with $400\;\mu m^2$. The diffusion coefficient, secretion rate and degradation rate of the chemoattractant were set equal to those used in our previous work \cite{Merks2008}; note that the diffusion coefficient is set to a value lower than the one, e.g., reported for VEGF in watery conditions ($10^{-11}\;\mathrm{m^2/s}$; see Ref.~\cite{Gamba2003}) because of its binding to ECM proteins \cite{KohnLuque:2013fw}. In absence of detailed experimental data on endothelial cell cell-cell and cell-ECM adhesive forces, cell stiffness, and the chemotactic response, for the corresponding parameters we used the values from Ref.~\cite{Merks2008}; the exact values of these parameters do not qualitatively affect the results of the model, and have modest quantitative impact; for a detailed sensitivity analysis see Refs.~\cite{Merks2008,Boas:2015td}.

\begin{table}[htbp]
  \caption{Parameter values for the angiogenesis and tip cell selection model. Underlined parameters are varied in the screen for tip cell behavior} 
  \begin{tabular}{|l|l|l|}
  \hline
  \textbf{Symbol}  & \textbf{description} & \textbf{value}\\
  \hline
  \underline{$\mu(\text{tip})$},$\mu(\text{stalk})$ & cell motility &  50 \\
   \underline{$J(\text{tip,stalk})$}, $J(\text{tip,tip})$, $J(\text{stalk,stalk})$ & cell-cell adhesion &40 \\
  \underline{$J(\text{tip,ECM})$}, \underline{$J(\text{stalk,ECM})$} & cell-ECM adhesion & 20 \\
  $A(\text{tip})$, $A(\text{stalk})$ & target area & 400 $\mathrm{\mu m^2}$ \\
  \underline{$\lambda(\text{tip})$}, $\lambda(\text{stalk})$ & elasticity parameter &  25\\
  \underline{$\chi(\text{tip})$}, $\chi(\text{stalk})$ & chemoattractant sensitivity &  500  \\
  \underline{$\alpha(\text{tip})$}, $\alpha(\text{stalk})$ & chemoattractant secretion rate &  $10^{-3}\;\text{s}^{-1}$ \\
  $\epsilon(\text{ECM})$ & chemoattractant decay rate in ECM & $10^{-3}\;\:\text{s}^{-1}$  \\
  $\epsilon(\text{tip})$, $\epsilon(\text{stalk})$ & chemoattractant decay rate below cells & $0\;\mathrm{{s}^{-1}}$\\
  D &  chemoattractant diffusion coefficient & $10^{-13}\;\text{m}^2\text{s}^{-1}$  \\
  \hline
  \end{tabular}
  \label{tab:par}
\end{table}

\subsection*{Computational screening for putative tip cell behavior}
We set up a screen for differences in the parameters of tip cells and stalk cells that affect the outcome of the model. In particular, we looked for parameters for which the tip cells lead sprouts in such a way that it affects the network morphology. In the angiogenesis model, a fraction 
($F_\text{tip}$) of the endothelial cells is assumed to be the ``tip cell'', $\tau(\sigma)=\text{tip}$, and the remaining fraction 
$1-F_\text{tip}$ is set to  $\tau(\sigma)=\text{stalk}$. We assigned the nominal parameters shown in Table ~\ref{tab:par} to both ``tip cells'' ($\tau(\sigma)=\text{tip}$) and ``stalk cells'' ($\tau(\sigma)=\text{stalk}$). We varied the underlined parameters in Table ~\ref{tab:par} to change the behavior of ``tip cells'' and ran the simulation for 10 000 time steps for a series of tip cell fractions and a series of parameters. The behavior of ``stalk cells'' was fixed because the nominal parameters, which were thoroughly studied in our previous work \cite{Merks2008}, are based on \emph{in vitro} experiments in which no tip cells were observed.

To keep this initial analysis computationally feasible, we tested only one parameter at a time instead of searching through the complete parameter space (see Ref.~\cite{Boas:2015td} for more systematic parameter study of the initial, single-cell-type model, based on a SOBOL-analysis). Also, in this initial screening we have limited the analysis to parameters that we could possibly associate directly with differentially expressed genes in tip and stalk cells. For this reason, we have omitted cell size differences, and we fixed the tip-tip cell adhesion strength. Fig~\ref{fig:parsweep}\textbf{B} illustrates a typical range of morphologies, or {\em morphospace}, that we obtained in this way. We analyzed the position of tip cells in each morphology (Fig~\ref{fig:parsweep}\textbf{C}) and 
analyzed the morphology of the vascular network in function of the tip cell fraction,  $F_\text{tip}$. 

To evaluate whether tip cells occupy sprout tips, we simulated the model with a tip cell fraction of $F_\text{tip}=0.2$, in accordance with published observations: 11.9\% in a HUVEC monolayer \cite{Siemerink2012} and $\sim$30\% in the growing of the retinal vasculature \cite{Hellstrom2007}. Because we assume that tip cell fate is strongly inhibited in a monolayer and tip cells are overexpressed in the growing front,  we set the tip cell fraction at 20\%, which is roughly the average of the two. 
At the end of each simulation we detected sprouts with tip cells on the tip using an automated method, as detailed in the Materials and Methods section. We then counted the percentage of sprouts with at least one tip cell at the sprout tip.  If more sprout tips were occupied by a tip cell than in the control experiment with identical tip and stalk cells, the parameter values were retained for further analysis. 

\begin{figure}[!htbp]
  \centering
\includegraphics{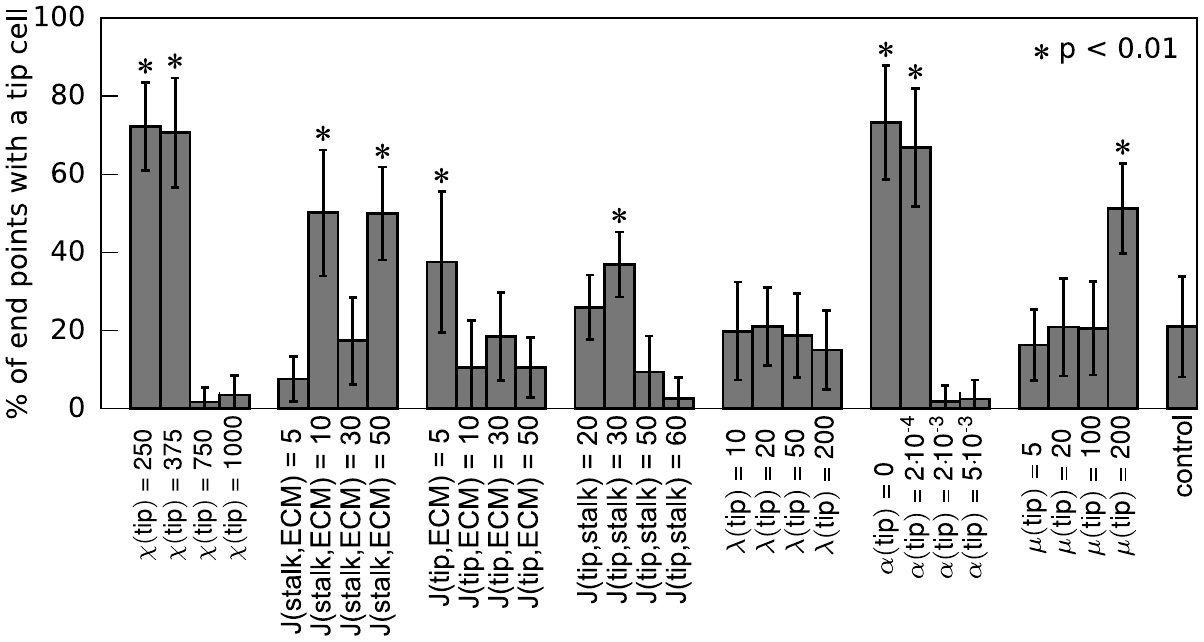}
  \caption{\textbf{Differences in cell properties can enable cells of one type to occupy sprout tips.} The percentage of sprout tips occupied by at least one tip cell was calculated at 10 000 MCS and averaged over 50 simulations (error bars depict the standard deviation). In each simulation 20\% of the cells were predefined as tip cells. For each simulation one tip cell parameter was changed, except for the control experiment where the nominal parameters were used for both tip and stalk cells. p-values were obtained with a one sided Welch's t-test for the null hypothesis that the number of tip cells at the sprout tips is not larger than in the control simulation.}
\label{fig:parsweep_tips}
\end{figure}

Fig~\ref{fig:parsweep_tips} shows the percentage of sprout tips occupied by one or more tip cells for all parameters tested. More sprouts are occupied by tip cells that: (a) are less sensitive to the autocrine chemoattractant than stalk cells ($\chi(\text{tip}) < \chi(\text{stalk})$), (b) adhere more strongly to the ECM than stalk cells ($J(\text{tip,ECM}) < J(\text{stalk,ECM})$), (c) adhere stronger to stalk cells than stalk cells to stalk cells ($J(\text{tip,stalk})<J(\text{stalk,stalk})$), (d) secrete the chemoattractant at a lower rate than stalk cells ($\alpha(\text{tip}) < \alpha(\text{stalk})$), or (e) have a higher active motility than stalk cells ($\mu(\text{tip}) > \mu(\text{stalk})$). 
For the parameters associated with cell-cell and cell-ECM adhesion, we observed a non-monotonic trend in Fig~\ref{fig:parsweep_tips}. A slight change in an adhesion parameter would affect the relative positions of tip and stalk cells, whereas a larger change can completely change the morphology of the network.  For example, if the tip cells adhere slightly more strongly to the ECM than the stalk cells, the tip cells tend to be pushed to the sprout tip (\nameref{S1_Fig}\textbf{A}). The stalk cells surround the tip cells if they adhere much more strongly to the ECM than the tip cells do (\nameref{S1_Fig}\textbf{B}), an effect that differential chemotaxis counteracts. In these simulations, the tip cells tended to cluster together. Because tip cells do not cluster together \cite{Gerhardt2003a}, we excluded reduced stalk-ECM adhesion from further analysis. 

Out of the cell behaviors that turned out to make cells move to the sprout tips, we next selected cell behaviors that also affect network morphology. We  quantified network morphology using two measures. The \emph{compactness}, $C=A_\text{cluster}/A_\text{hull}$ is the ratio of the area of the largest cluster of connected cells, $A_\text{cluster}$, and the area of the convex hull enclosing the connected cluster, $A_\text{hull}$ \cite{Merks2008}. It approaches $C=1$ for a disk and tends to $C\rightarrow 0$ for a sparse network. We also counted the number of ``gaps'' in the network, or lacunae, $N_\mathrm{lacunae}$. For details see the Materials and Methods section. 

\begin{figure}[!htbp]
  \centering
  \includegraphics[width=\textwidth]{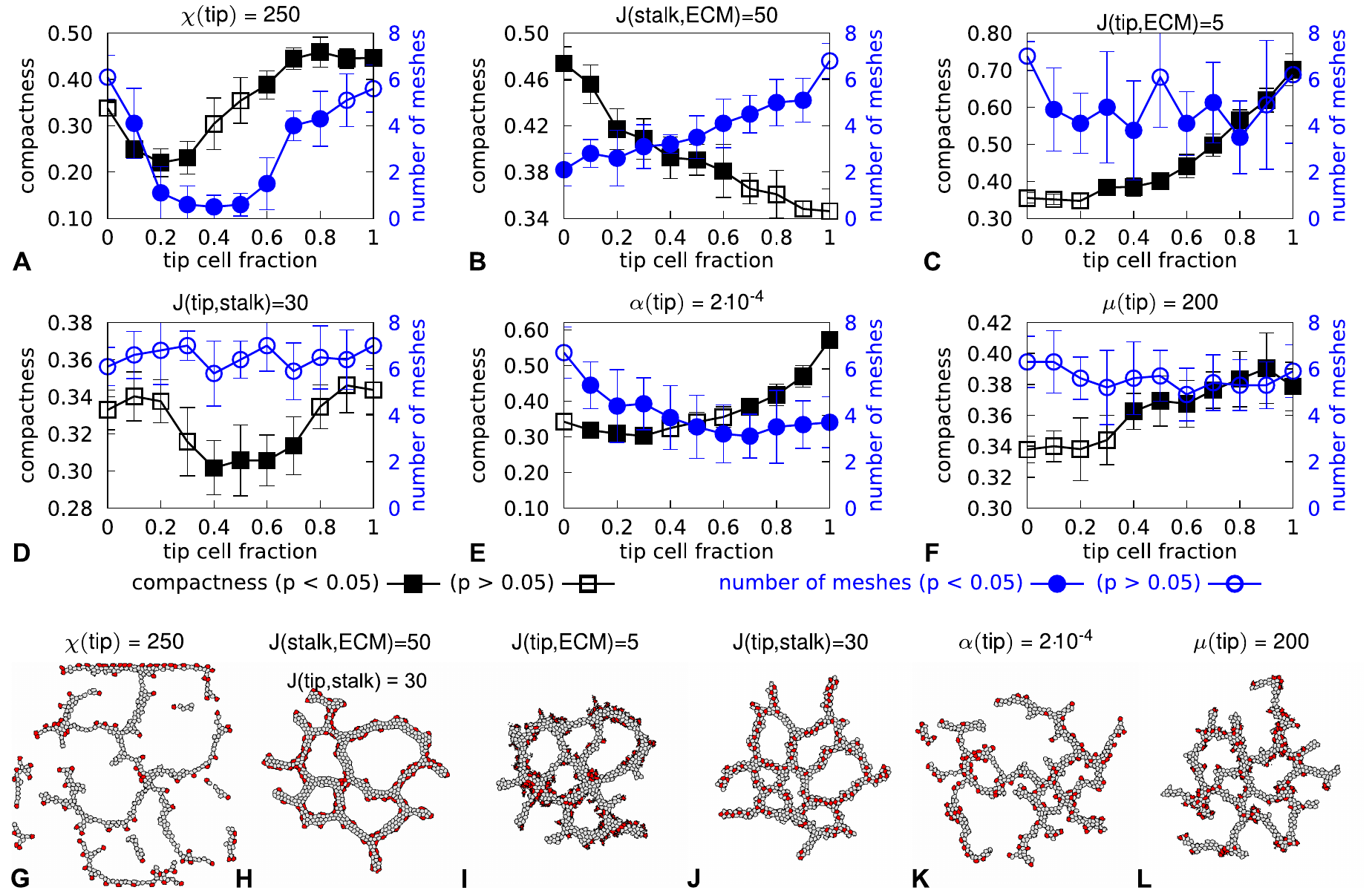}
  \caption{\textbf{Effects of different tip and stalk cell properties on network morphology.} \textbf{A}-\textbf{F} Trends of compactness (black rectangles) and number of lacunae (blue circles) calculated with the morphologies at 10 000 MCS. For each data point 10 morphologies were analyzed and the error bars represent the standard deviation. p-values were obtained with a Welch's t-test for the null hypothesis that the mean of the sample is identical to that of a reference with the nominal parameters listed in Table 1. For \textbf{B} this reference is the data for tip cell fraction 1 and for all other graphs this is the data for tip cell fraction 0. \textbf{G}-\textbf{L} Morphologies after 10 000 MCS for each tested parameter value with $F_\text{tip}$ = 0.2.}
  \label{fig:parsweep_morph}
\end{figure}

Figs~\ref{fig:parsweep_morph}\textbf{A}-\textbf{F} takes a selection of the tip cell parameters identified in the previous section, and then plots the compactness $C$ (black curves) and the number of lacunae $N_\mathrm{lacunae}$ (blue curves) as a function of the tip cell fraction. The results for the remaining parameter value are shown in (\nameref{S1_Fig}). For each tip cell fraction tested, the outcome is then compared with simulations in which the tip cells where identical to the stalk cells (i.e., as in Fig~\ref{fig:parsweep}\textbf{A}). Closed symbols indicate a significant difference with the respective reference simulation (Welch's t-test, $p<0.05$, $n=10$). Tip cell parameters that affected network morphologies for at least half of the tip cell fractions tested were kept for further analysis. 

The screening selected three ways in which tip cells could differ from stalk cells to change network morphology: reduced chemoattractant sensitivity ($\chi(\text{tip}) < \chi(\text{stalk})$; see Fig~\ref{fig:parsweep_morph}\textbf{A}), reduced chemoattractant secretion by tip cells ($\alpha(\text{tip}) < \alpha(\text{stalk})$; see Fig~\ref{fig:parsweep_morph}\textbf{E}), and increased tip-ECM adhesion ($J(\text{stalk,ECM}) > J(\text{tip,ECM})$; see Figs~\ref{fig:parsweep_morph}\textbf{B}-\textbf{C}). It turned out that increased ECM adhesion by tip cells was best modeled  by reducing the adhesion of stalk cells with the ECM instead ($J(\text{stalk,ECM})$), because for $J(\text{tip,ECM})=5$ (Fig~\ref{fig:parsweep_morph}\textbf{C}) networks could not form with too many tip cells (see \nameref{S1_Video}). 

The results of the screening held for the other parameter values tested (\nameref{S1_Fig}) with two exceptions: (1) the networks disintegrated if tip cells did not respond sufficiently strongly to the chemoattractant ($\chi(\mathrm{tip})<100$ (\nameref{S2_Fig}\textbf{J})), and (2) the tip cells spread out over the stalk cells to cover the whole network  for $J(\text{stalk,ECM}>70)$ (\nameref{S2_Fig}\textbf{K}). Also, the conclusions were confirmed in a screening relative to three additional nominal parameter sets  (\nameref{S3_Fig} and \nameref{S4_Fig}).

Altogether, the computational screening presented in this section identified three tip cell parameters that affect tip cell position in the sprout and the morphology of the networks formed in our computational model: reduced secretion of the chemoattractant, reduced sensitivity to the chemoattractant, and increased tip-ECM adhesion. It is possible, however, that these effects are due to spatial or temporal averaging of tip and stalk cell parameters, not due to interaction of two different cell types. The next section will introduce a control for such effects.

\subsection*{Comparison with control model selects  ``reduced chemoattractant sensitivity'' scenario for further analysis}

The computational screening highlighted three tip cell parameters that affected both the position of tip cells in the sprouts and the morphology of the networks: (1) increased tip-cell ECM adhesion, (2) reduced chemoattractant secretion by tip cells, and (3) reduced chemoattractant sensitivity of tip cells. Because it was unsure whether these effects were due to (a) the differential cell behavior of tip and stalk cells, or (b) due to temporal or spatial averaging of the parameters differentially assigned to tip and stalk cells, we compared the results against a control model that had only one cell type with ``averaged'' parameters: $P(\text{cell}) = (1-F_\text{tip})\cdot P(\text{stalk})+F_\text{tip}\cdot P(\text{tip})$, with $P(\text{tip})$ the tip cell parameter value and $P(\text{stalk})$ the stalk cell parameter value. 

\begin{figure}[!htbp]
  \centering
\includegraphics[width=\textwidth]{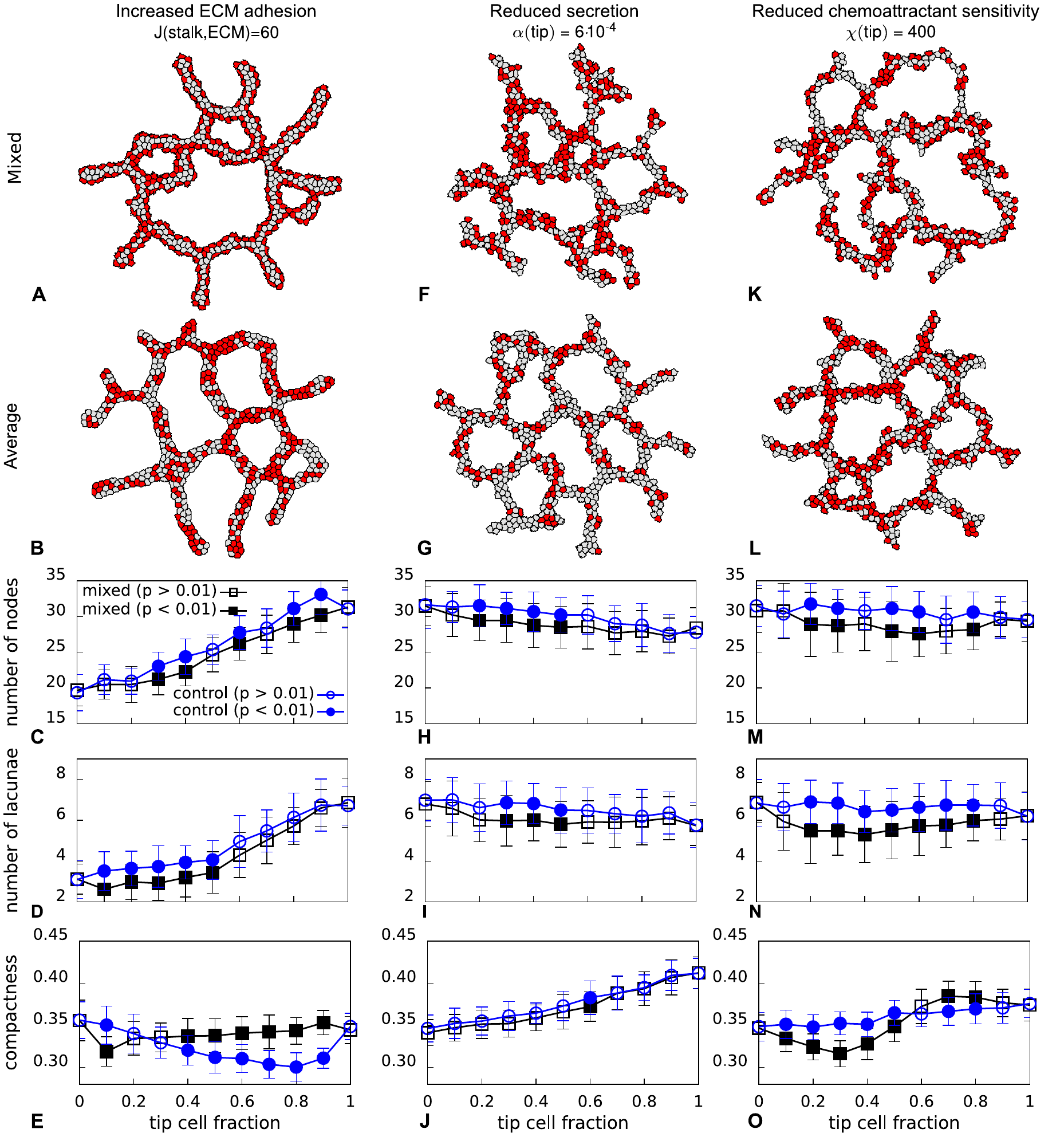}
  \caption{\textbf{Comparison of networks formed with mixed cells and cells with average properties.} \textbf{A}, \textbf{F}, and \textbf{K} morphologies for mixed tip (red) and stalk (gray) cells ($F_\text{tip} = 0.5$). \textbf{B}, \textbf{G}, and \textbf{L} morphologies for averaged cells ($F_\text{tip} = 0.5$). \textbf{C}-\textbf{E}, \textbf{H}-\textbf{J}, and \textbf{M}-\textbf{O} morphometrics for a range of tip cell fractions for both the control and mixed model. The morphometrics were calculated for 50 simulations at 10 000 MCS (error bars represent the standard deviation). p-values were obtained with a Welch's t-test for the null hypothesis that the mean of mixed model and the control model are identical.}
  \label{fig:tcpropmetric}
\end{figure}

For each of the three parameters identified in the first step of the computational screening, we compared the morphologies formed in the control model after 10000 MCS with the morphologies formed in the original model with mixed cell types (Fig~\ref{fig:tcpropmetric}).  Figs~\ref{fig:tcpropmetric}\textbf{A,F, and K} show example configurations formed  in the original model, in comparison with example configuration formed in the corresponding ``averaged'' model (Figs~\ref{fig:tcpropmetric}\textbf{B,G, and L}).  In the ``mixed'' model the tip cells (red) tend to move to the periphery of the branches, in contrast to the ``averaged'' model in which all cells have the same parameter values. 

We next tested if networks formed in the ``mixed'' model differed from those formed in the corresponding ``averaged'' model for tip cell fractions ranging from 0 (no tip cells) to 1 (only tip cells). Although the measures differed for individual morphometrics and tip cell fractions in all three scenarios (Figs~\ref{fig:tcpropmetric}\textbf{C-E, H-J, M-O}), only in the model where tip cells had reduced chemoattractant sensitivity all morphometrics differed significantly for practically all tip cell fractions tested (Figs~\ref{fig:tcpropmetric}\textbf{M}-\textbf{O}).  The analysis was repeated for three additional parameter values per scenario  (\nameref{S5_Fig}); although in all three scenarios the morphometrics differed between the ``mixed'' and ``averaged'' models for a number of tip cell fractions, only in the ``reduced chemoattractant sensitivity'' scenario the differential behavior of tip and stalk cells consistently affected the morphometrics. We thus retained only this model for further analysis.

\subsection*{Heterogeneous chemoattractant sensitivity increases direction persistence of migrating tip and stalk cell pairs}
\begin{figure}[!htbp]
  \centering
\includegraphics[width=.9\textwidth]{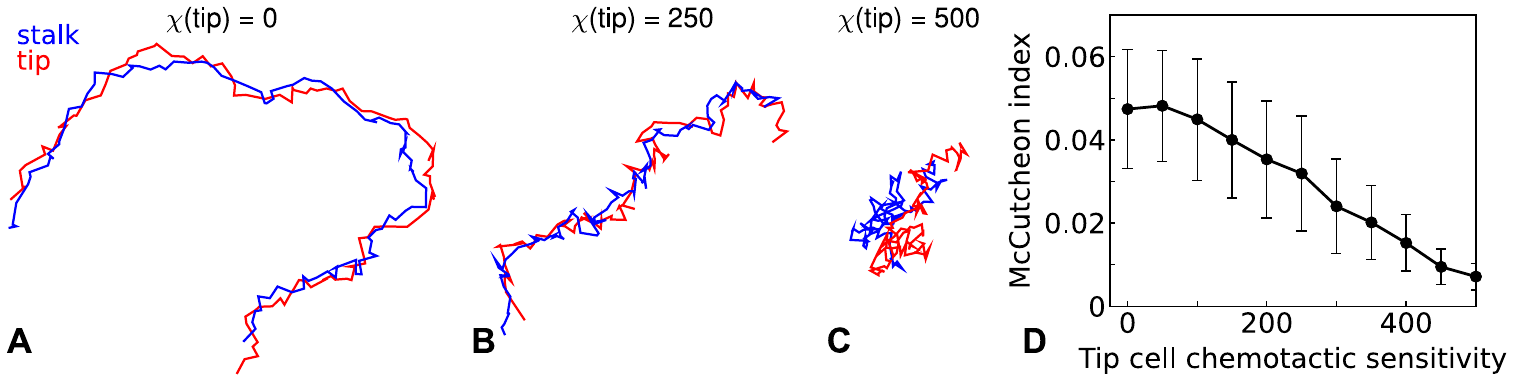}
  \caption{Migrating cell pairs consisting of a stalk cell and a tip cell with a reduced chemoattractant sensitivity. \textbf{A}-\textbf{C} Trajectories of the tip and stalk cell during 10 000 MCS with $\chi(\text{stalk})=500$ and respectively $\chi(\text{tip}) = 0$, $\chi(\text{tip}) = 250$, and $\chi(\text{tip}) = 5000$. \textbf{D} McCutcheon index as a function of the tip cell chemoattractant sensitivity. The values were averaged over 100 simulations and error bars depict the standard deviation.}
  \label{fig:cellpair}
\end{figure}

The parameter screening indicated that tip cells that are less sensitive to the chemoattractant than stalk cells tend to move to the front of the sprouts, affecting in this way the network morphology. To better understand now such differential chemoattractant sensitivity can affect angiogenic sprouting, we analyzed the migration of a cell pair consisting of one tip cell and one stalk cell. As shown in Figs~\ref{fig:cellpair}\textbf{A}-\textbf{C}, cell pairs with a large difference in the chemoattractant sensitivity migrated much further than cell pairs with a smaller or no difference in chemoattractant sensitivity. To quantify this observation, we used the McCutcheon index \cite{McCutcheon1946}, which is the ratio of the distance between the initial and final position, and the total path length. As shown in Fig~\ref{fig:cellpair}\textbf{D}, the McCutcheon index decreases as the tip cell's chemoattractant sensitivity approaches that of the stalk cell. Indicating that a strong difference in chemotaxis causes the cell pair to move along a straighter path. These results suggest that, in a self-generated gradient, heterogeneous chemoattractant sensitivity improves migration speed and persistence. In the context of angiogenesis, this effect speeds sprouting and sprout elongation.

\subsection*{Local tip cell selection regularizes network morphology}

In the parameter screenings presented in the previous sections, to a first approximation we assumed that a subpopulation of endothelial cells are ``predetermined''  to become tip cells, e.g., due to prior expression of CD34 \cite{Siemerink2012}.  It is likely, however, that tip cell fate is continuously ``re-evaluated'' in a Dll4-Notch-VEGFR2 signaling loop \cite{Hellstrom2007,Suchting2007a,Siekmann2007,Lobov2007a}. Tip cells express Dll4 on their cell membranes \cite{Claxton2004}, which binds to the Notch receptor on adjacent cell membranes. This leads to the release of the Notch intracellular domain (NICD), activating the stalk cell phenotype \cite{Hellstrom2007,Siekmann2007}. Via this lateral inhibition mechanism, cells adjacent to tip cells tend to differentiate into stalk cells. To simulate such ``dynamic tip cell selection'', a simplified genetic regulatory network (GRN) model of Dll4-Notch signaling was added to each simulated cell, as described in detail in the  Materials and Methods. Briefly,  the level of NICD in each cell is a function of the amount of Dll4 expressed in adjacent cells, weighed according to the proportion of the cell membrane shared with each adjacent cells. If the concentration of NICD $N(\sigma)$ of a tip cell $\sigma$ exceeds a threshold $N(\sigma)>\Theta_\text{NICD}$, the cell cross-differentiates into a stalk cell; conversely, if in a stalk cell $N(\sigma)\leq\Theta_\text{NICD}$ it differentiates into a tip cell \cite{Hellstrom2007,Suchting2007a,Siekmann2007}. 

 \begin{figure}[!htbp]
   \centering
   \includegraphics[width=\textwidth]{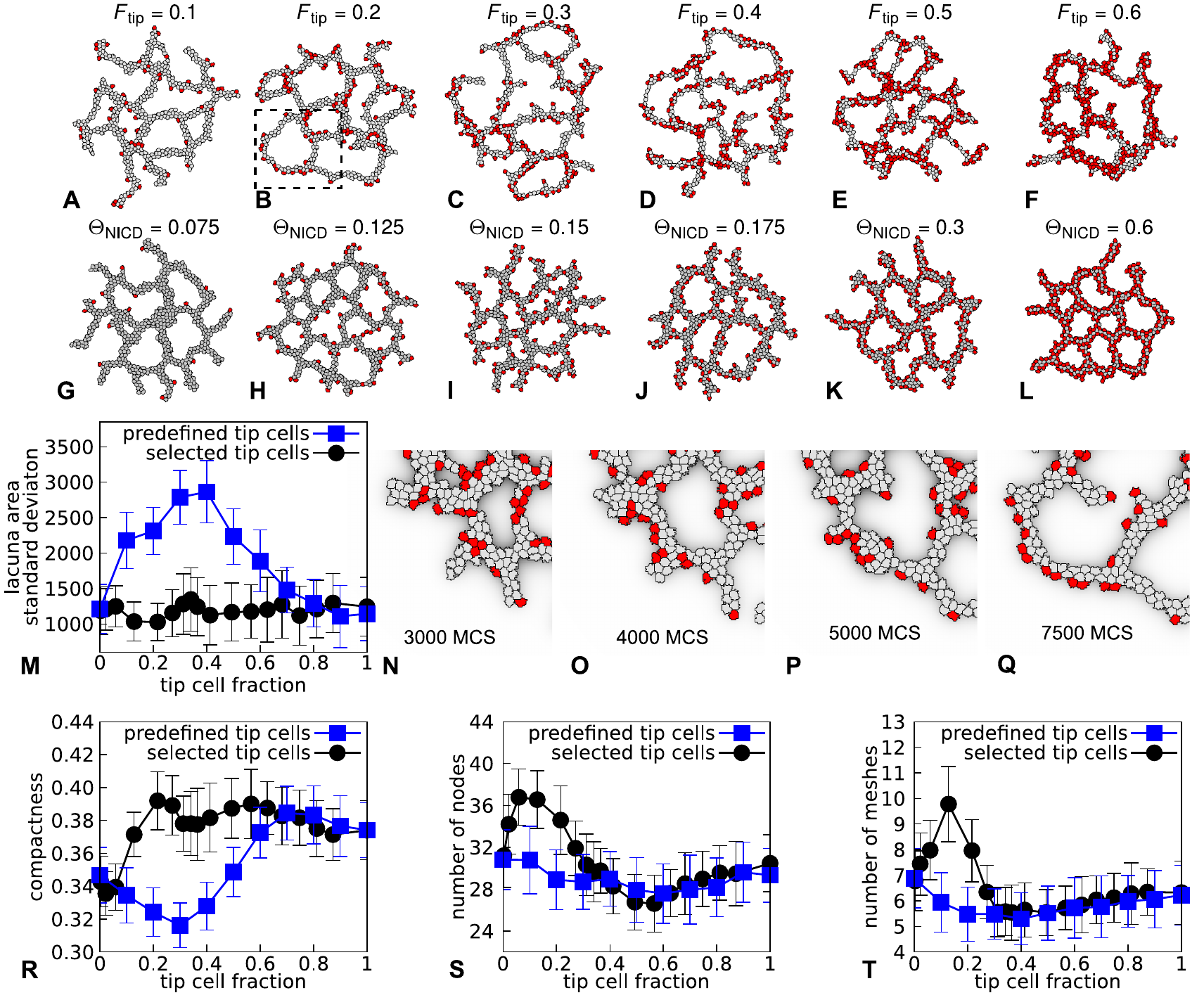}
   \caption{\textbf{Effects of tip cell selection on network formation.} \textbf{A}-\textbf{F} Networks formed with varying fractions of predefined tip cells ($F_\text{tip}$) with $\chi(\text{tip})=400$ at 10 000 MCS. \textbf{G}-\textbf{L} Networks formed with the tip cell selection model for varying NICD thresholds ($\Theta_\text{NICD}$) at 10 000 MCS. \textbf{M} Standard deviation of lacuna area in a network after 10 000 MCS. \textbf{N}-\textbf{Q} Close up of the evolution of a network with 20\% predefined tip cells (marked area in \textbf{B}). \textbf{R}-\textbf{T} Comparison of the morphometrics for networks formed with predefined and selected tip cells with reduced chemoattractant sensitivity ($\chi(\text{tip})=400$) and network at 10 000 MCS. For the simulations with tip cell selection, the average tip cell fraction was calculated for each NICD threshold. For all plots (\textbf{M} and \textbf{R}-\textbf{T}) the values were averaged over 50 simulations and error bars depict the standard deviation.}
   \label{fig:ptcVSdtc}
 \end{figure}

Fig~\ref{fig:ptcVSdtc} shows the behavior of the initial ``static model'' (Figs~\ref{fig:ptcVSdtc}\textbf{A}-\textbf{F}) in comparison with the ``dynamic tip cell selection'' model (Figs~\ref{fig:ptcVSdtc}\textbf{G}-\textbf{L}). In the dynamic model the tip cell fraction was set using the values of $\Theta_\text{NICD}$, such that the exact tip cell fractions depended on the local configurations.  In comparison with the initial, ``static'' model (Figs~\ref{fig:ptcVSdtc}\textbf{A}-\textbf{F}), the model with ``dynamic'' selection (Figs~\ref{fig:ptcVSdtc}\textbf{G}-\textbf{L} and \nameref{S3_Video}) seems to form more compact and regular networks. To quantify this difference in network regularity, we determined the variation of the areas of the lacunae of the networks at the final time step of a simulation. Fig~\ref{fig:ptcVSdtc}\textbf{M} shows this measurement averaged over 50 simulations for a range of tip cell fractions. Lacunae in networks formed from mixtures of stalk cells and 10\% to 60\% ``static'' tip cells have more variable sizes than lacunae in networks formed by the `dynamic tip cell' model. 

To further analyze how dynamic tip cell selection regularized network morphologies in our model, we studied in detail how tip cells contributed to network formation in the ``static'' and ``dynamic'' tip cell models. Figs~\ref{fig:ptcVSdtc}\textbf{N}-\textbf{Q} shows the evolution of a part of a network formed with 20\% ``static tip cells''. At first, some tip cells locate at sprout tips and others are located adjacent to or within the branches (Fig~\ref{fig:ptcVSdtc}\textbf{N}). The chemoattractant gradually accumulates ``under'' the branches, with a curvature effect producing slightly higher concentrations at the side of the lacunae. This attracts the stalk cells (Fig~\ref{fig:ptcVSdtc}\textbf{O}), ``squeezing'' the tip cells out of the branch and away from the lacuna, due to their reduced chemoattractant sensitivity (Fig~\ref{fig:ptcVSdtc}\textbf{P} and \nameref{S2_Video}). The resulting layered configuration with tip cells at the outer rim drives a drift away from the lacuna  (Fig~\ref{fig:ptcVSdtc}\textbf{Q}): Due to their stronger chemoattractant sensitivity, the stalk cells attempt to move to the center of the configuration, pushing the tip cells away, thus leading to directional migration driven by the mechanism outlined in the previous section (see also Ref.~\cite{Dona2013}).

In the ``dynamic tip cell selection'' mechanism, the persistent migration will be confined to the sprout tips. The model thus suggests that tip cells could assist in producing a local, self-generated gradient mechanism that directs the migration of sprouts, a mechanism that requires tip cells to differentiate only at sprout tips. For tip cells to  ``drag'' just the sprouts,  only a limited number of tip cells must be present in the network.  To test this idea, we compared network morphologies for the ``dynamic'' and the ``static'' tip cell models for a range of tip cell fractions (Figs~\ref{fig:ptcVSdtc}\textbf{R}-\textbf{T}). Indeed, the network morphologies were practically identical for high tip cell fractions, whereas they differed significantly for all three morphometrics for tip cell fractions between 0.1 and 0.3: In the dynamic selection model the networks become more disperse (Fig~\ref{fig:ptcVSdtc}\textbf{R}) and formed more branches (Fig~\ref{fig:ptcVSdtc}\textbf{S}) and lacunae (Fig~\ref{fig:ptcVSdtc}\textbf{T}) than in the ``static'' model. 

\begin{figure}[!htbp]
  \centering
  \includegraphics[width=\textwidth]{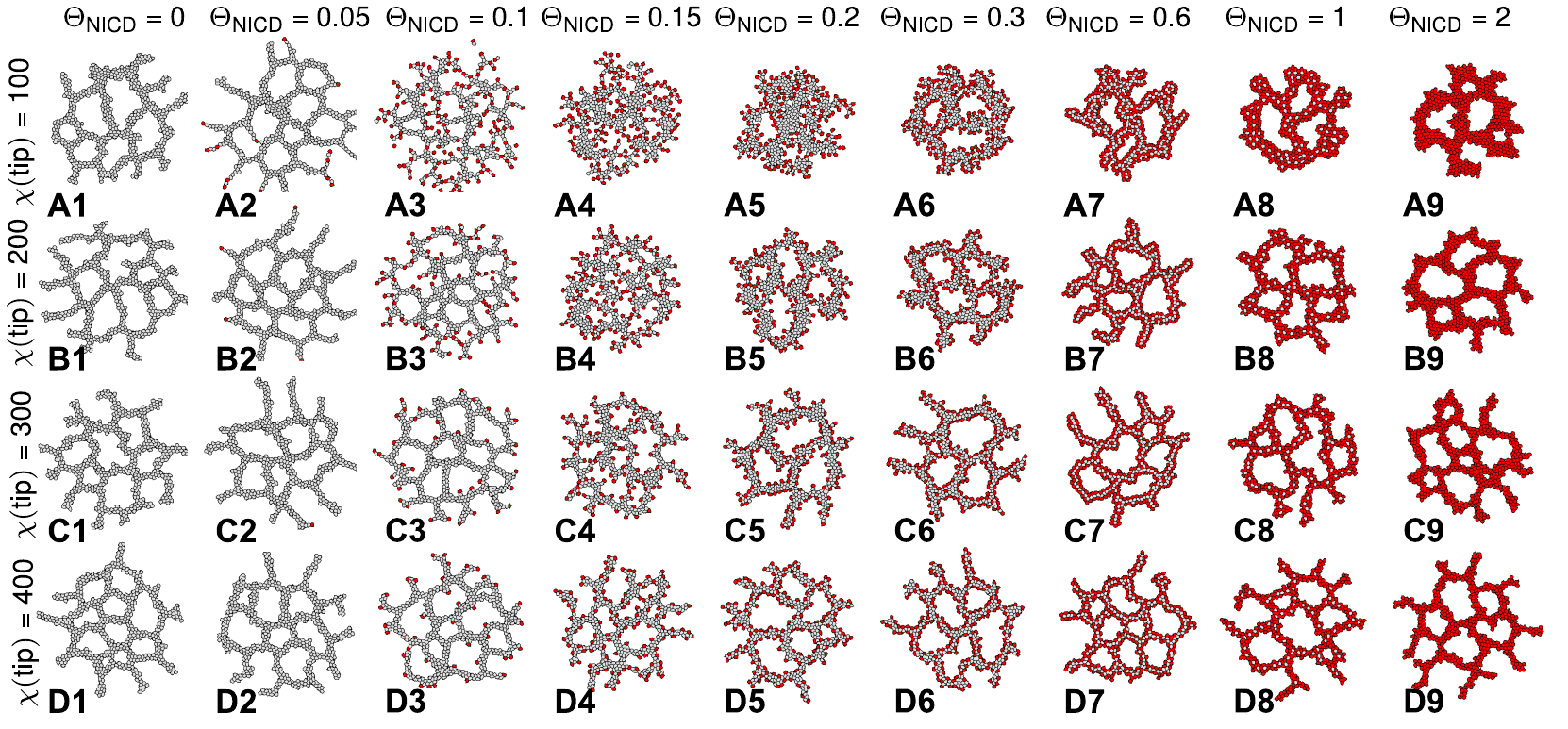}
  \caption{\textbf{Effects of reducing tip cell chemoattractant sensitivity for varying NICD thresholds.} Morphospace of the final morphologies (10 000 MCS) with varying tip cell chemoattractant sensitivities ($\chi(\text{tip})$) and NICD thresholds ($\Theta_\text{NICD}$).}
  \label{fig:dtc_Ct}
\end{figure}

To validate the ``dynamic'' tip cell model, we compared the effect of the tip cell fraction on network morphology with published experimental observations. The \emph{in vivo}, mouse retinal angiogenesis model is a good and widely used model for tip/stalk cell interactions during angiogenesis \cite{Ridgway2006a,Suchting2007a,Lobov2007a,Hellstrom2007,Arima2011,Jakobsson2010,Benedito2012,Benedito2009a,Bentley2014}. Networks formed with an increased abundance of tip cells become more dense and form a larger number of branches \cite{Hellstrom2007,Suchting2007a,Lobov2007a,Ridgway2006a} than wild type networks. Our computational model is consistent with this trend for tip cell fractions between 0 and up to around 0.2 (Figs~\ref{fig:ptcVSdtc}\textbf{R}-\textbf{T}), but for tip cell fractions $>0.2$ the vascular morphologies become less branched (Figs~\ref{fig:ptcVSdtc}\textbf{S}-\textbf{T}). To investigate in more detail to what extent our model is consistent with these experimental observations, we tested the effect of the tip cell fraction in the `dynamic' tip cell selection model in more detail. In particular we were interested in how the difference in chemoattractant sensitivity between tip and stalk cells affected network morphology. Fig~\ref{fig:dtc_Ct} shows the effect of the NICD threshold (increasing the NICD threshold is comparable to inhibiting Dll4 expression or Notch signaling, and hence controls the tip cell fraction) for a range of tip cell chemoattractant sensitivities. When the difference in the chemoattractant sensitivity between tip and stalk cells is relatively small ($\chi(\text{tip}) \geq 300)$), increasing the NICD threshold results in the formation of denser network with fewer lacunea. In contrast, when the difference in chemoattractant sensitivity between tip and stalk cells is larger ($\chi(\text{tip} \leq 200$), there exists an intermediate state in which the networks are both compact and have a large number of branch points (Figs~\ref{fig:dtc_Ct}\textbf{A4} and \textbf{B4}). This intermediate state resembles the dense, highly connected networks that are observed when tip cells are abundant in the mouse retina \cite{Hellstrom2007,Suchting2007a,Lobov2007a,Ridgway2006a}. Thus, when the difference in the chemoattractant sensitivity of tip and stalk cells is sufficiently large, the model can reproduce both normal angiogenesis and the excessive angiogenic branching observed for an abundance of tip cells \cite{Hellstrom2007}.

\subsection*{Survey for chemoattractant receptors reduced in tip cells suggests Apelin as candidate}

The comparative, computational model analysis of the role of tip cells in angiogenesis, predicted that--among the models tested--a model where tip cells show reduced sensitivity to an autocrine chemoattractant best matches tip cell phenomenology: The tip cells lead the sprouts, and facilitate the formation of vascular networks of regular morphology for tip cell fractions of up to around 0.2. Could a chemoattractant with these, or very similar properties be involved in vascular development? To answer this question,  we evaluated four comparative studies of gene expression in tip and stalk cells \cite{Harrington2008a,DelToro2010,Strasser2010,Siemerink2012}. These studies identified three receptors involved in endothelial chemotaxis that were differentially expressed in tip cells and stalk cells: VEGFR2, CXCR4, and APJ.
VEGFR2 is upregulated in tip cells \cite{Harrington2008a,Siemerink2012,Bentley2014}. VEGFR2 is a receptor for the chemoattractant VEGF that is secreted by hypoxic tissue \cite{Ferrara2004}. Whether or not VEGF is secreted at sufficiently high levels to act as an autocrine chemoattractant between endothelial cells has been under debate \cite{Seghezzi1998,Merks2008,Franco2011}, with the emerging being that it is most likely a long-range guidance cue of angiogenic sprouts secreted by hypoxic tissues (\hspace{1sp}\cite{Gerhardt2003a}; reviewed in Ref.~\cite{Geudens2011}).  The chemokine CXCL12 and its receptor CXCR4 \cite{Gupta1998} are both upregulated in tip cells \cite{Harrington2008a,Siemerink2012,Strasser2010}, suggesting that tip cells would have higher, not lower sensitivity to CXCL12 signaling than stalk cells. Interestingly, CXCL12 and CXCR4 are key components of a self-generated gradient mechanism for directional tissue migration in the lateral line primordium mechanisms \cite{Dona2013}. Because of the key role of CXCL12/CXCR4 in angiogenesis (see, e.g., \cite{Salcedo2003}) it is therefore tempting to speculate that CXCL12/CXCR4 may be part of a similar, self-generated gradient mechanism during angiogenesis.

However, because CXCL12 expression is upregulated in tip cells relative to stalk cells, not downregulated, we will focus here on a third receptor/ligand pair differentially expressed in tip and stalk cells: APJ and Apelin. APJ is a receptor for the endothelial chemoattractant Apelin \cite{Tatemoto1998,Kasai2004,Kalin2007} that is secreted by endothelial cells \cite{Kasai2004,Kalin2007}. Apelin expression is upregulated in tip cells  \cite{Siemerink2012,DelToro2010,Strasser2010}, whereas its receptor APJ is not detected in tip cells \cite{DelToro2010}.  Thus the expression pattern of Apelin and its receptor APJ fits with our model prediction: Apelin is an endothelial chemoattractant that is secreted by endothelial cells and tip cells are less responsive to Apelin than stalk cells. In our model the chemoattractant is secreted at the same rate by tip and stalk cells, whereas Apelin is preferentially expressed in tip cells. The next section will therefore add preferential secretion of Apelin by tip cells to the model, and test if and how this changes the predictions of our model.

\subsection*{Model refinement to  mimic role of Apelin/APJ more closely}

\begin{figure}[!htbp]
  \centering
  \includegraphics{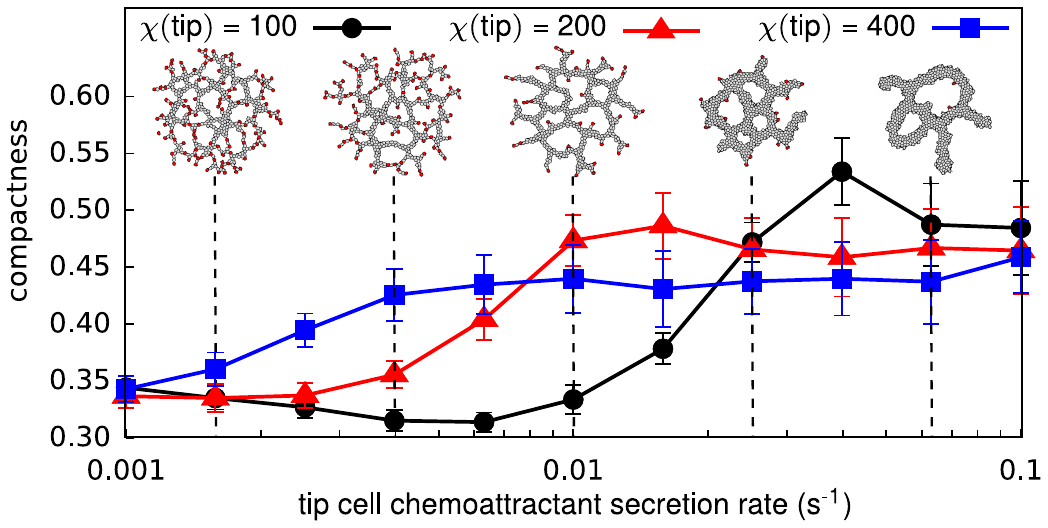}
  \caption{\textbf{Effects of increasing tip cell Apelin secretion rate for varying levels of tip cell chemotaxis.} Compactness of the final network (10 000 MCS) with the morphologies for $\chi(\text{tip})=100$ for tip cell Apelin secretion rates of $\alpha(\text{tip})=1.6\cdot 10^{-3}$, $\alpha(\text{tip})=4.0\cdot 10^{-3}$, $\alpha(\text{tip})=1\cdot 10^{-2}$,$\alpha(\text{tip})=2.5\cdot 10^{-2}$, and $\alpha(\text{tip})=6.3\cdot 10^{-2}$ as insets. Except for $\alpha(\text{tip})$, all parameters have the values listed in Table~\ref{tab:par}. Data points show average values for $n=50$ simulations with error bars giving the standard deviation.}
  \label{fig:dtc_apelin}
\end{figure}

The computational analyses outlined in the previous sections suggest that Apelin and its receptor APJ might act as an autocrine chemoattractant in the way predicted by our model: Both stalk cells and tip cells secrete Apelin and APJ \cite{Kasai2004,Kalin2007} and the tip cells do not express the APJ receptor \cite{DelToro2010}. Gene expression analyses \cite{Siemerink2012,DelToro2010} also suggest that tip cells secrete Apelin at a higher rate than stalk cells. We therefore tested if the simulation results still held if we changed the model assumptions  accordingly: In addition to a reduced chemoattractant sensitivity in tip cells ($\chi(\text{tip})=100$), we assumed tip cells secrete chemoattractant at a higher rate than stalk cells: $\alpha(\text{tip})>\alpha(\text{stalk})$.  Although the absence of APJ expression in tip cells suggests that tip cells are insensitive to the chemoattractant, $\chi(\text{tip})=0$, to reflect the phenomenological observation that endothelial cells are attracted to one another, we set $\chi(\text{stalk})>\chi(\text{tip})>0$. 
Such intercellular attraction could, e.g.,  be mediated by cell-cell adhesion,  by alternative chemoattractant-receptor pairs (e.g., CXCR4-CXCL12 \cite{Salvucci2002}), or by means of mechanical endothelial cell interactions via the extracellular matrix\cite{Reinhart-King2008}. Fig~\ref{fig:dtc_apelin} shows how the Apelin secretion rate in tip cells ($\alpha(\text{tip})$) affects the morphology of the vascular networks formed in our model, as expressed by the compactness. For tip cell secretion rates of up to around $\alpha(\text{tip})=0.01$ the model behavior does not change. The networks became more compact and exhibit thicker branches for tip cell chemoattractant secretion rates of $\alpha(\text{tip})>0.01$. This result does not agree with the observation that Apelin promotes vascular outgrowth \cite{Kasai2004,Cox2006}. The increased compactness for $\alpha(\text{tip})>0.01$ is a model artifact: stalk cells were so strongly attracted to tip cells that they engulfed the tip cells and thereby inhibited the tip cell phenotype. A similar increase in compactness and branch thickness is observed in a model where tip cells are not sensitive to the chemoattractant (\nameref{S6_Fig}), which indicates that a too large Apelin secretion rate of tip cells destabilizes sprout elongation.
Altogether, these results suggest that, if the Apelin secretion rate of tip cells does not become more than ten times larger than that of stalk cells, our model produces similar results independent of the tip cell secretion rate of Apelin.

\subsection*{Apelin or APJ silencing inhibits sprouting \emph{in vitro} and \emph{in silico}}

\begin{figure}[!htbp]
  \centering
  \includegraphics[width=\textwidth]{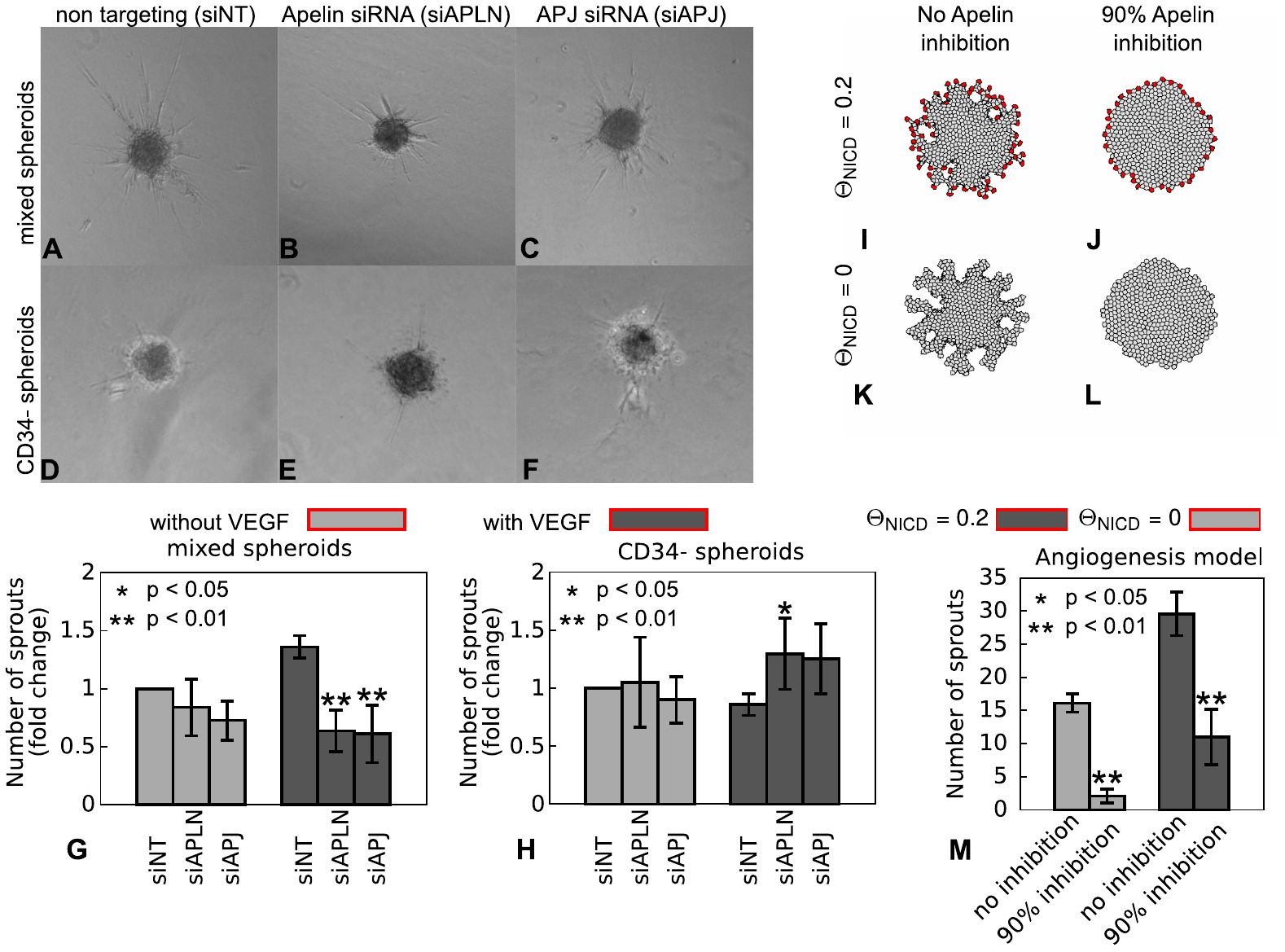}
 \caption{\textbf{Effects of Apelin or APJ silencing in spheroid sprouting assays.} \textbf{A}-\textbf{F} Microscopy images of the WT and CD34- spheroids in VEGF-enriched collage after 24 hours. \textbf{G}-\textbf{H} Number of sprouts, relative to siNT treatment, after 24 hours for spheroids with mixed cells and CD34- spheroids. These metrics are the mean of the normalized, average number of sprouts of each replicate with the error bars depicting standard deviation. The * denotes $p < 0.05$, see Materials and Methods for details of the normalization and statistical analysis. \textbf{I}-\textbf{L} Example morphologies formed in the computational angiogenesis model (750 MCS); \textbf{(I-J)} model including tip cells ($\theta_\mathrm{NICD}=0.2$, in absence (\textbf{I})  and in presence (\textbf{J}) of chemoattractant inhibition; \textbf{(K-L)} model with reduced tip cell number  ($\theta_\mathrm{NICD}=0$) in presence (\textbf{K}) and in absence (\textbf{M}) of chemoattractant inhibition. \textbf{M} Number of sprouts after 750 MCS for $n=20$ simulations; error bars show the standard deviation; asterisks denote $p < 0.05$ for p-values obtained with Welch's t-test in comparison with controls (no inhibition).}
  \label{fig:silexp}
\end{figure}

Previous studies have shown that Apelin promotes angiogenesis of retinal endothelial cells seeded on Matrigel \cite{Kasai2004}, as well as in \emph{in vivo} systems such as the mouse retina, {\em Xenopus} embryo, and chick chorioallantoic membrane \cite{Cox2006}.  Furthermore, \emph{in vivo} inhibition of Apelin or APJ reduced sprouting in {\em Xenopus} embryos \cite{Cox2006}, zebrafish \cite{Eyries2008}, and the mouse retina \cite{Kasai2008,DelToro2010}.  To assess the relation between tip-stalk cell interaction and Apelin signaling, we inhibited Apelin signaling in an \emph{in vitro} model of angiogenic sprouting in which the fraction of CD34- (``stalk'') cells could be controlled. Spheroids of immortalized human microvascular endothelial cells (HMEC-1s) were embedded in collagen gels and in collagen enriched with VEGF.  After culturing the spheroids for 24 hours at 37 degrees Celsius under 5\% $\mathrm{CO_2}$, the cultures were photographed (Figs~\ref{fig:silexp}A-F and \nameref{S2_Dataset}). The spheroids did not form network structures within the culturing time, whereas the computational model simulated both angiogenenic sprouting and subsequent vascular plexus formation (Fig~\ref{fig:parsweep}A). In order to assess the effect of Apelin and APJ silencing on sprouting in the \emph{in vitro} and \emph{in silico} models, we assess the morphologies formed by the \emph{in silico} model after 750 MCS. For each model the degree of sprouting was assessed by counting the number of sprouts using the semi-automated image analysis software ImageJ. We compared sprouting in a ``mixed'' spheroid of HMEC-1s with a population enriched in ``stalk cells'', i.e., a population of CD34- HMEC-1s sorted using FACS.  To inhibit Apelin signaling, the spheroids were treated with an siRNA silencing translation of Apelin (siAPLN) or of its receptor (siAPJ).  

Figs~\ref{fig:silexp}\textbf{A}-\textbf{F} and \textbf{K}-\textbf{L} show how the number of sprouts per spheroid changes, relative to the treatment with non-translating siRNA (siNT), due to the silencing RNA treatments. To determine significance, ANOVA was performed on each data set, one for the ``mixed'' spheroids and one the ``stalk cell'' spheroids, and followed up by pairwise comparisons using Tukey's range test (see Materials and Methods for detail). Relative to a control model with non-translating siRNA (siNT), ``mixed'' spheroids in VEGF enriched collagen formed fewer sprouts (Figs~\ref{fig:silexp} \textbf{A}-\textbf{C} and \textbf{G}, and \nameref{S5_Fig}) when treated with siAPJ or siAPLN. Interestingly, when the collagen gels are not enriched with VEGF, siAPJ or siAPLN did not significantly affect the number of sprouts. Since VEGF can induce tip cell fate \cite{Benedito2009a,Caolo2010}, this may suggest that without VEGF there are too few tip cells present to observe the effects of inhibiting Apelin signaling. 
In ``stalk cell'' spheroids siRNA treatments interfering with Apelin treatments slightly improved sprouting in some replicates and had no clear effect in others (\nameref{S5_Fig}). Thus these results suggest that Apelin signaling requires a mix of sufficient CD34+ (``tip'') and CD34- (``stalk'') cells, in support of our hypothesis that differential chemotaxis of stalk and tip cells to Apelin drives the sprout forward.

We next asked if the observed reduction of sprouting associated with inhibition of Apelin-signaling also occurred in the computational model. To mimic application of siAPLN in the computational model, we reduced the secretion of the chemoattractant both in tip and stalk cells to $\alpha(\mathrm{tip})=10^{-3}$ and $\alpha(\mathrm{stalk})=10^{-4}$. To mimic wild-type spheroids we used  $\Theta_\text{NICD}=0.2$, which yields a mix of CD34+ and CD34- cells. To mimic spheroids enriched in stalk cells, we reduced the NICD-levels to $\Theta_\text{NICD} = 0$ in which case all endothelial cells became stalk cells. Figs~\ref{fig:silexp}\textbf{I}-\textbf{L}  and \nameref{S4_Video} show how the model responds to the inhibition of Apelin-signaling, showing reduced sprouting after inhibiting the chemoattractant. To quantify these observations, we repeated the simulations ten times for 750 MCS. We converted the resulting images to gray scale images  (see \nameref{S8_Fig}) and counted the number of sprouts using ImageJ, thus using the same quantification procedure as that used for the \emph{in vitro} cultures. In both the \emph{in silico} ``wild type'' spheroids ($\Theta_\text{NICD}=0.2$) and in the \emph{in silico} ``stalk cell'' spheroids ( $\Theta_\text{NICD} = 0$), inhibition of Apelin-signaling reduced sprouting. However, the simulations did not reproduce the experimental observation that in ``stalk cell'' spheroids silencing of Apelin signaling had little effect in absence of VEGF and slightly promoted sprouting in VEGF-treated CD34- cultures.

\section*{Discussion}
In this work we asked how and by what mechanisms tip cells can participate in angiogenic sprouting. We employed a suitable computational model of angiogenic network formation \cite{Merks2008}, which was extended with tip and stalk cell differentiation. In the extended model, the behavior of tip and stalk cells could be varied independently by changing the model parameters. Instead of testing preconceived hypotheses on tip and stalk cell behavior, we took a ``reversed approach'' in which we could rapidly compare series of alternative parameter settings, each representing different tip cell behavior: We systematically searched for parameters that led tip cells to occupy the sprouts tips, and that changed the morphology of the angiogenic networks relative to a nominal set of simulations in which tip and stalk cells have identical behavior. We studied two cases, reflecting the two extremes in the range of known molecular mechanisms regulating tip and stalk cell differentiation. In the first case, we assumed that endothelial cells are differentiated stably between a tip and stalk cell phenotype within the characteristic time scale of angiogenic development (approximately 24 to 48 hours). In the second case, we assumed a much more rapidly-acting lateral inhibition mechanism, mediated by Dll4 and Notch. Here endothelial cells can switch back and forth between tip and stalk cell fate at time scales of the same order of cell motility. Our analysis showed that in a model driven by contact-inhibited chemotaxis to a growth factor secreted by endothelial cells, tip cells that respond less to the chemoattractant move to the tips of the sprouts and speed up sprout extension. Under the same conditions, more regular and more dense networks formed if endothelial cells switched between tip and stalk cell fate due to lateral inhibition. This limits tip cells  to growing sprouts; due to their stronger chemoattractant sensitivity the stalk cells push the tip cells forwards leading to faster sprout extension in a mechanism reminiscent of a ``self-generated gradient mechanism'' \cite{Dona2013}.

We next asked if a growth factor with the predicted properties is involved in angiogenic sprouting. To this end we looked for matching, differential gene expression patterns in published data sets of gene expression in tip and stalk cells. In particular the Apelin-APJ ligand-receptor pair turned out to be a promising candidate:  Apelin is a chemoattractant for endothelial cells that is secreted by endothelial cells and the receptor APJ is only detected in stalk cells.  In agreement with our  simulations, \emph{in vitro} experiments on endothelial spheroids showed that inhibition of Apelin or its receptor APJ reduced \emph{in vitro} spheroid sprouting. Thus the reversed bottom-up simulation approach employed in this study helped identify a candidate molecule mediating the interaction between tip and stalk cells during angiogenesis. 

Our approach was inspired by a prior study that used a computational model to identify what cell behavior changed when endothelial cells were treated with certain growth factors \cite{Long2012}. This study used an agent-based, 3D model of angiogenesis in which sprouts extend from a spheroid. With a genetic algorithm the parameters for which the model reproduces experimental results are derived. In this way Long \emph{et al.} \cite{Long2012} could hypothesize what changes in cell behavior the growth factors caused and successfully derived how certain growth factors affect cell behavior in 3D sprouting assays. Here, we used a similar approach to study what behavior makes tip cells lead sprouts and affect network formation, using high-throughput parameter studies instead of objective optimization approaches.  Tip-stalk cell interactions have been studied before with several hypothesis-driven models where specific behavior was assigned to the tip cells based on experimental observations, and tip cells were either defined as the leading cell \cite{Milde2008a,Bauer2007a,Bauer2009,Artel2011,Mehdizadeh2013,Jackson2010} or tip cell selection was modeled such that the tip cell could only differentiate at the sprout tip \cite{Qutub2009d,Travasso2011,Kleinstreuer2013}. These models have been used to study how extracellular matrix (ECM) density \cite{Milde2008a}, ECM degradation \cite{Milde2008a}, ECM inhomogeneity \cite{Bauer2007a,Bauer2009}, a porous scaffold \cite{Artel2011,Mehdizadeh2013}, cell migration and proliferation \cite{Jackson2010,Qutub2009d}, tip cell chemotaxis \cite{Travasso2011} and toxins \cite{Kleinstreuer2013} affect sprouting and angiogenesis. Thus these studies asked how a specific hypothesis of tip cell behavior and tip cell position affected the other mechanisms and observables in the simulation. Our approach aims to develop new models for the interaction between tip and stalk cells that can reproduce biological observation.  These new hypotheses can be further refined in hypothesis-driven model studies, as we do here, e.g., in Fig~\ref{fig:dtc_apelin}.

In order to make this ``reversed' approach possible, we have simplified the underlying genetic regulatory networks responsible for tip-stalk cell differentiation. These molecular networks, in particular Dll4-Notch signaling, have been modeled in detail by Bentley \emph{et al.} \cite{Bentley2008,Bentley2009}. Their model describes a strand of endothelial cells, and was used to study how lateral inhibition via Dll4-Notch signaling in interaction with VEGF signaling participates in tip cell selection. With this model Bentley and coworkers predicted that the shape of the VEGF gradient determines the rate of tip cell selection, and that for very high levels of VEGF the intracellular levels of Dll4 and VEGFR2 oscillate. Based on their experimental observations that tip cells migrate within a sprout, cell movement has been added to the model by allowing cells to switch positions along the sprout \cite{Jakobsson2010}. Bentley and coworkers reproduced tip cell migration in the sprout and showed that the VEGFR2 levels in a cell determine the chance of an endothelial cell to become a tip cell. The migration of tip cells in a sprouts was further studied using a model that included a cell migration model \cite{Bentley2014}. Bentley and coworkers \cite{Bentley2014} thus showed that the differences in VE-cadherin expression between tip and stalk cells could cause tip cell migration to the sprout tip. Altogether, these models gave useful insights in the role of Dll4-Notch signaling and VEGF signaling in tip cell selection in a growing sprout. Here, instead of focusing at single sprouts, we focused on the scale of a vascular network. By combining a tip cell selection model with a cell based model of angiogenesis, we showed that tip cell selection can aid the development of dense networks by limiting the destabilizing effects of tip cells.

The model prediction that tip cells respond less to a chemoattractant secreted by all endothelial cells fits with the expression pattern of the chemoattractant Apelin, which is secreted by all endothelial cells and of which the receptor is not detect in tip cells. Previous studies indicated that Apelin induces angiogenesis \emph{in vitro} \cite{Kasai2004,Kalin2007}.  Apelin-APJ signaling is necessary for vascular development in \emph{in vivo} systems such as in the mouse retina \cite{Kasai2008}, frog embryo \cite{Cox2006,Kalin2007}, and chicken chorioallantoic membrane \cite{Cox2006}. Furthermore, high levels of vascularization in human glioblastoma are correlated with high expression levels of Apelin and APJ \cite{Kalin2007}. Based on these observations Apelin is considered to be a pro-angiogenic factor. Similar to other pro-angiogenic factors such as VEGF \cite{Bautch2012a}, Apelin is expressed near areas where blood vessels develop and Apelin expression is induced by hypoxia \cite{Eyries2008}. The pro-angiogenic role of Apelin is linked to its role as a chemoattractant \cite{Cox2006,Eyries2008} and mitogenic factor \cite{Cox2006,Eyries2008}. However, the role of Apelin in proliferation may be disputed because Apelin did not promote proliferation in a series of sprouting assays with human umbilical vein endothelial cells, human umbilical arterial endothelial cells, and human dermal microvascular endothelial cells \cite{Kalin2007}. Our models propose a scenario where Apelin can promote angiogenesis as an autocrine chemoattractant, in contrast to the previous studies where the source of Apelin was external. Such a mechanism would fit with the observation that the Apelin receptor APJ is only expressed in stalk cells. 

Inhibition of sprouting is manifested as a decrease in the number of sprouts. As mentioned previously, Apelin may promote proliferation, and thus inhibition of Apelin signaling may results in a reduced proliferation rate. A reduced proliferation rate could result in a reduced sprout length, but, a reduced number of sprouts is an unlikely effect of a decreased proliferation rate. This indicates that the mechanism that drives sprouting is affected by the inhibition of Apelin signaling. However, whereas in the model inhibition of Apelin signaling inhibits sprouting for all tested cases, in the experimental assays the effects of Apelin or APJ inhibition depended on the fraction of tip cells and the environment. In mixed spheroids, Apelin and APJ inhibition reduced sprouting in spheroids embedded in VEGF-enriched collagen. In CD34- spheroids, i.e., spheroids enriched with stalk cells, Apelin or APJ inhibition had no effect in plain collagen and slightly enhanced sprouting in a VEGF rich environment. This suggests that, in a VEGF rich environment, Apelin-APJ signaling inhibits sprouting by stalk cells. VEGF has been shown to induce tip cell fate \cite{Benedito2009a,Caolo2010}, as well as APJ expression \cite{Kidoya2008,Hara2013}. However, it remains unclear how the combination of a VEGF rich environment and Apelin signaling could inhibit sprouting and therefore further experiments studying the interaction between VEGF and Apelin signaling in vascular sprouting are needed. Further \emph{in vitro} experiments are also needed to study the effects of Apelin signaling on network formation, that follows the initial sprouting phase. Our model predicts that inhibition of Apelin signaling would also block network formation (see \nameref{S4_Video}). However, because the 3D sprouting assay does not mimic vascular network formation, this prediction could not be verified experimentally.

The importance of VEGF in our validation experiments suggests that we cannot ignore VEGF in our tip cell selection model. As mentioned above, VEGF may interact with Apelin-APJ signaling. Furthermore, VEGF \cite{Ferrara2004} and Apelin \cite{Cox2006,Eyries2008} are both involved in endothelial cell proliferation. Besides the link between VEGF and Apelin, VEGF is also involved in tip cell selection.
Dll4-Notch signaling and VEGF signaling interact directly in two ways. First, Dll4 is upregulated by signaling between VEGF and VEGF receptor 2 (VEGFR2) \cite{Benedito2009a,Caolo2010}. Second, Dll4-Notch signaling downregulates VEGFR2 \cite{Suchting2007a,Williams2006a,Harrington2008a,Taylor2002} and upregulates VEGF receptor 1 (VEGFR1) \cite{Harrington2008a,Funahashi2010}, which acts as a decoy receptor for VEGF \cite{Park1994}. Because \emph{in vivo} VEGF acts as an external guidance cue for angiogenesis, the interplay between VEGF signaling and Dll4-Notch signaling could promote tip cell selection in the growing sprouts. The expression levels of VEGFR2 also directly reduce adhesion between cells because VEGFR2-VEGF binding causes endocytosis of VE-cadherin \cite{Gavard2006}. This reduced adhesion may enable cells with high VEGFR2 levels, such as tip cells, to migrate to the sprout tip \cite{Bentley2014}. 
Because of this complex interplay between between cell behavior and  Dll4, Notch, VEGF, and the VEGF receptors, future studies will replace the simplified tip cell selection model for a tip cell selection model with explicit levels of Dll4, Notch, VEGF, VEGFR1 and VEGFR2, and link those levels directly to tip and stalk cell behaviors. Furthermore, future studies should include explicit levels of Apelin and APJ to study if and how VEGF-induced Apelin secretion affects network formation. Such an extended model will provide more insight into  how the interaction between stalk cell proliferation \cite{Gerhardt2003a,Kasai2013}, ECM association of VEGF \cite{Ruhrberg2002}, and pericyte recruitment and interaction \cite{Ribatti2011,Kasai2013}, which all have been linked to Apelin signaling and/or VEGF signaling, affects angiogenesis. 

\section*{Materials and Methods}

\subsection*{Cellular Potts model}
In the cellular Potts model \cite{Graner1992,Glazier1993a} cells are represented on a finite box $\Lambda\subset\mathbb{Z}^{2}$ within a regular square lattice. Each lattice site $\vec{x}\in\Lambda$ represents a $2\mu m \,\times\,2\mu m$ portion of a cell or the extracellular matrix. They are associated with a cell identifier $\sigma \in \mathbb{Z}^{\{+,0\}}$. Lattice sites with $\sigma=0$ represent the extracellular matrix (ECM) and groups of lattice sites with the same $\sigma > 0$ represent one cell. Each cell $\sigma$ has a cell type $\tau(\sigma) \in \{\text{ECM,tip,stalk}\}$. The balance of adhesive, propulsive and compressive forces that cells apply onto one another is described using a Hamiltonian,
\begin{equation}
 H = \underbrace{\sum _{(\vec{x},\vec{x}^\prime)} J(\tau,\tau^\prime) (1-\delta(\sigma,\sigma^\prime))}_\text{cell adhesion} + \underbrace{\vphantom{\sum _{(\vec{x},\vec{x}^\prime)}}\sum_\sigma \lambda(\tau(\sigma)) \left( a(\sigma)-A(\tau(\sigma)) \right )^2}_\text{area constraint}, \label{eq:ham}
\end{equation}
with $(\vec{x},\vec{x}^\prime)$ a set of adjacent lattice sites, $\tau = \tau(\sigma(\vec{x}))$ and $\tau^\prime = \tau(\sigma(\vec{x}^\prime))$, $\sigma = \sigma(\vec{x})$ and $\sigma^\prime = \sigma(\vec{x}^\prime)$, $J(\tau,\tau^\prime)$ the contact energy, the Kronecker delta: $\delta(x,y) = \left \{ 1, x = y; 0, x \neq y \right\}$, the elasticity parameter $\lambda(\tau)$, and the target area $A(\tau)$. To mimic random pseudopod extensions the CPM repeatedly attempts to copy the state $\sigma(\vec{x})$ of a randomly chosen lattice site $\vec{x}$, into an adjacent lattice site $\vec{x}^\prime$ selected at random among the eight nearest and next-nearest neighbors of $\vec{x}$. The copy attempt is accepted with probability,
\begin{equation}
  \label{eq:paccept}
  p_\text{accept}(\Delta H) = \left \{ \begin{array}{ll}
    1 & \text{if }\Delta H \leq 0; \\
    e^{\frac{-\Delta H}{f(\tau,\tau^\prime)}} & \text{if }\Delta H > 0;
    \end{array} \right .
\end{equation}
with
\begin{equation}
  f(\vec{x},\vec{x}^\prime) = \left \{\begin{array}{ll}
    \min(\mu(\tau),\mu(\tau^\prime)) & \text{if }\sigma > 0 \text{ and } \sigma^\prime > 0; \\
    \max(\mu(\tau),\mu(\tau^\prime)) & \text{otherwise}.
  \end{array} \right .
\end{equation}
Here is $\mu(\tau)$ is the cell motility and $\tau=\tau(\sigma(\vec{x}))$ and $\tau\prime=\tau(\sigma(\vec{x}\prime))$ are shorthand notations. One Monte Carlo step (MCS)---the unit time step of the CPM---consists of $|\Lambda|$ random copy attempts; i.e., in one MCS as many copy attempts are performed as there are lattice sites in the simulation box.

The endothelial cells secrete a chemoattractant at rate $\alpha(\tau)$ that diffuses and decays in the ECM,
\begin{equation}
 \frac{\partial c(\vec{x},t)}{\partial t} = D \nabla^2 c(\vec{x},t) + \alpha({\tau(\sigma(\vec{x})}))(1-\delta(\sigma(\vec{x}),0)) - \varepsilon \delta(\sigma(\vec{x}),0)c(\vec{x},t), \label{eq:diffusion}
\end{equation}
with $c$ the chemoattractant concentration, $D$ the diffusion coefficient, and $\varepsilon$ the decay rate. After each MCS equation \ref{eq:diffusion} is solved numerically with a forward Euler scheme using 15 steps of $\Delta t=2\text{s}$ and a lattice spacing coinciding with the cellular Potts lattice of $\Delta x=2\mu\text{m}$ with absorbing boundary conditions ($c=0$ at the boundaries of $\Lambda$); thus one MCS corresponds with 30 seconds. Chemotaxis is modeled with a  gradient dependent term in the change of the Hamiltonian \cite{Savill1996} associated to a copy attempt from $\vec{x}$ to $\vec{x}^\prime$:
\begin{equation}
 \Delta H _\text{chemotaxis} = -\chi(\tau,\tau^\prime) \left(\frac{c(\vec{x}^\prime)}{1+sc(\vec{x}^\prime)}-\frac{c(\vec{x})}{1+sc(\vec{x})} \right), \label{eq:chemotaxis}
\end{equation}
with $\chi(\tau,\tau^\prime)$ the chemoattractant sensitivity of a cell of type $\tau$ towards a cell of type $\tau^\prime$ and vice versa, and $s$ the receptor saturation. In the angiogenesis model we assumed that chemotaxis only occurs at cell-ECM interfaces (contact-inhibited chemotaxis; see \cite{Merks2008} for detail); hence we set $\chi(\tau)=0$ if $\tau\ne\text{ECM}$ and $\tau^\prime\ne\text{ECM}$. For the remaining, non-zero chemoattractant sensitivities we use the shorthand notation $\chi(\tau)$.

\subsection*{Tip cell selection model}
The differentiation between tip and stalk cells is regulated by a simplified tip and stalk cell selection model. The model is based on lateral inhibition via Dll4-Notch signaling: If Dll4 binds to Notch on a adjacent cell it causes the dissociation of Notch, resulting in the release of Notch intracellular domain (NICD) \cite{Bray2006a}. We assume that tip cells express Notch at a permanent level of $\mathcal{N}(\text{tip})$ and Delta at a level of $\mathcal{D}(\text{tip})$; stalk cells express Delta and Notch at permanent levels of $\mathcal{N}(\text{stalk})$ and $\mathcal{D}(\text{stalk})$. The level of NICD in a cell, $\mathcal{I}(\sigma)$, is given by,
\begin{equation}
 \mathcal{I}(\sigma) = \frac{\mathcal{N}(\tau(\sigma))}{a(\sigma)} \sum_{n \in \text{neighbors}}  \mathcal{D}(\tau(n)) L_{\sigma \cap n}, \label{eq:nicd}
\end{equation}
in which $\mathcal{N}(\tau)$ and $\mathcal{D}(\tau)$ are the levels of Notch and Delta in a cell of type $\tau$, and $L_{\sigma \cap n}$ is the length of the interface between cells $\sigma$ and $n$. To model differentiation between the stalk and tip cell type in response to the release of NICD \cite{Hellstrom2007,Siekmann2007} the cell type is a function of the cell's NICD level,
\begin{equation}
  \tau(\sigma) =
  \begin{cases}
    \text{tip} & \text{if } \mathcal{I}(\sigma) \leq \Theta_\text{NICD}; \\
    \text{stalk} & \text{if } \mathcal{I}(\sigma) > \Theta_\text{NICD},
  \end{cases}
\end{equation}
with $\Theta_\text{NICD}$ threshold representing the NICD-level above which the cell differentiates into a stalk cell. To prevent rapid cell type changes, we introduced a hysteresis effect by setting the Notch levels to: $\mathcal{N}(\text{tip})=0.3$ and $\mathcal{N}(\text{stalk})=0.5$. The Dll4 levels are set according to the experimental observation that tip cells express more membrane bound Dll4 than stalk cells \cite{Claxton2004}:  $\mathcal{D}(\text{tip})=4$ and $\mathcal{D}(\text{stalk})=1$.

\subsection*{Morphometrics}
To quantify the results of the sprouting simulations we calculated the compactness of the morphology and detect the lacunae, branch points and end points.
The compactness $C$ is defined as $C = A_\text{cell}/A_\text{hull}$, with $A_\text{cell}$ the total area of a set of cells and $A_\text{hull}$ the area of the convex hull around these cells. For the compactness we used the largest connected component of lattice sites with $\sigma > 0$. This connected component was obtained using a standard union-find with path compression \cite{Galler1964}. The convex hull around these lattice sites is the smallest convex polygon that contains all lattice sites which is obtained using the Graham scan algorithm \cite{Graham1972}.

Lacunae are defined as connected components of lattice sites with $\sigma(\vec{x})=0$ (ECM) completely surrounded by lattice sites with $\sigma(\vec{x}) > 0$. These areas are detected by applying the \emph{label} function of Mahotas on the binary image $\left\{\vec{x}\in\Lambda,1_{\sigma(\vec{x})=0}\right\}$, i.e., the image obtained if medium pixels are set to 1 and all other pixels are set to 0. The number of labels areas in this image is the number of lacuna, and the number of lattice sites in a labeled area is the area of a lacuna.

To identify the branch points and end points, the morphology is reduced to a single pixel morphological skeleton \cite{Guidolin2004}. For this, first the morphology is obtained as the binary image $\left\{\vec{x}\in\Lambda,1_{\sigma(\vec{x})>0}\right\}$. Rough edges are removed from the binary image by applying a morphological closing \cite{Dougherty2003} with a disk of radius 3. Then, 8 thinning steps are performed in which iteratively all points that are detected by a hit-and-miss operator are removed from the image \cite{Dougherty2003}. In the skeleton, pixels with more than two first order neighbors are branch points and pixels with only one first order neighbor are end points. The skeleton may contain superfluous nodes. Therefore, all sets of nodes that are within a radius of 10 lattice units are collected and replaced by a single node at: $\vec{n}_\text{merged} = \langle \vec{x} \rangle_{\{\vec{x}\in\text{nodes}:|\vec{n}-\vec{x}|<10\}}$. 

All morphological operations are performed using the Python libraries Mahotas \cite{Coelho2013} and Pymorph \cite{pymorph}. Mahotas implements standard morphological operations, except for the closing and thinning operations required for skeleton generation. For these we use Pymorph, that implements a more complete set of morphological operation than Mahotas. However, as it is implemented in pure Python it is computationally less efficient than Mahotas.

\subsection*{Tip cell detection}
\begin{figure}[!htbp]
  \centering
  \includegraphics[width=\textwidth]{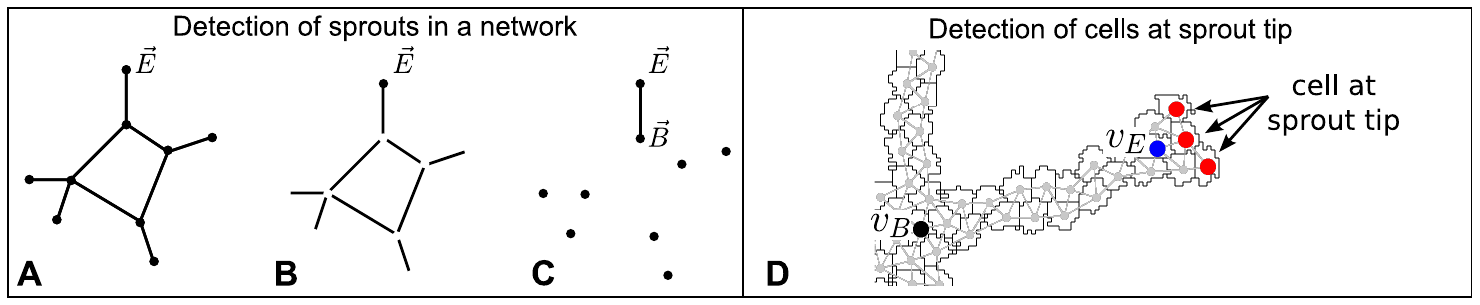}
  \caption{\textbf{Detection of cells at the tips of sprouts.} \textbf{A}-\textbf{C} detection of sprouts in a network. \textbf{A} Skeleton with branch points and end points. \textbf{B} Skeleton from which all nodes except $\vec{E}$ are removed. \textbf{C} The union of the nodes and the connected component in \textbf{B} that contains $\vec{E}$. The node that, in \textbf{C}, is part of the same connected component as $\vec{E}$ is the branch point $\vec{B}$. \textbf{D} detection of cells at the sprout tip (red vertices), which are farthest away from the branch point $v_B$ (black vertex).}
  \label{fig:tipdetection}
\end{figure}

Cells at the sprout tips were automatically detected in two steps:
(1) detection of the sprouts in the network; (2) detection of the cells on the sprout tip. For the first step, detecting sprouts, a sprout is defined as a connection between a branch point, $\vec{B}$, and an end point, $\vec{E}$. To find the branch point $\vec{B}$ that is connected to end point $\vec{E}$, all nodes, except $\vec{E}$, are removed from the morphological skeleton (Fig~\ref{fig:tipdetection}\textbf{B}). In the resulting image one part of the skeleton is still connected to $\vec{E}$, this is the branch. Then, all nodes are superimposed on the image with the branch (Fig~\ref{fig:tipdetection}\textbf{C}) and the node connected to $\vec{E}$ is the branch point $\vec{B}$. 
Next, we search for the cells at the tip of the sprout, which are the cells in the sprout furthest away from $\sigma_B$. To find these cells we use a graph representation of the morphology. In this graph, $G(v,r)$, each vertex $v$ represents a cell and vertices of neighboring cells share an edge (Fig~\ref{fig:tipdetection}\textbf{D}). Now, we calculate the shortest path between each vertex $v$ and the vertex belonging to the cell at the branch point $v_B$ using Dijkstra's algorithm \cite{Dijkstra1959}. Then, we iteratively search for vertices with the longest shortest path to $v_B$ starting at the vertex associated to $\sigma_E$ ($v_E$). To limit the search to the a single sprout, the search is stopped when $v_B$ is reached. When the search is finished, the node or nodes with the longest shortest path to $v_B$ represent the cells or cells that are at the sprout tip.

\subsection*{Model implementation and parameter sweeps}

The simulations were implemented using the cellular Potts modeling framework \emph{CompuCell3D} \cite{Swat2012} which
can be obtained from \url{http://www.compucell3D.org}. The simulation script is deposited in \nameref{S1_File}. File \nameref{S1_File} also includes two extensions to \emph{CompuCell3D}, called steppables, which we developed for the simulations presented in this paper. Steppable \emph{RandomBlobInitializer} is used to initialize the simulations with a blob of cells, and steppable \emph{TCS} contains the tip cell selection model. To efficiently set up, run and analyze large parameters sweeps including the ones presented in this paper, we have developed a pipeline to set up, run, and analyze large numbers of simulations of cell-based models on parallel hardware using software like \emph{CompuCell3D}, described in detail elsewhere \cite{Palm2014}. Briefly, the pipeline automatically generate simulation scripts for a list of parameters values, run the simulations on a cluster, and analyze the results using the morphometric methods described in sections Morphometrics and Tip Cell Detection.

\subsection*{\emph{In vitro} sprouting assay}
Immortalized human dermal endothelial cells (HMEC-1s) \cite{vanHengel1997} were cultured in 2\% gelatin-coated culture flask at 37 $^\circ$C under 5\% $\mathrm{CO_2}$ with a M199 medium (Gibco, Grand Island, NY, USA) supplemented with 10\% foetal calf serum (Biowhittaker, Walkersvillle, MD, USA), 5\% human serum and 1\% Penicillin-streptomycin-glutamine (Gibco). The HMEC-1 cells used in this study were a kind gift of Prof. Dr. P. Hordijk (Sanquin, Amsterdam, the Netherlands) and were derived from Ref.~\cite{vanHengel1997}. Cell suspensions were obtained from the cultures by TrypLE (Gibco) treatment of adherent endothelial cell monolayers. After the cells were extracted from the culture they were seeded in methylcellulose (Sigma-Aldrich) containing medium to allow spheroid formation \cite{Korff1998}. After 18 hours, the spheroids were embedded in a collagen gel containing human serum. In the period that these experiments were performed, the lab had to change collagen gels because of availability issues. Therefore, the following three gels were used: Purecol bovine collagen (Nutacon, Leimuiden, the Netherlands), Nutacon bovine collagen (Nutacon, Leimuiden, the Netherlands), and Cultrex rat collagen I (R\&D Systems, Abingdon, United Kingdom). The gels may be supplemented with VEGF-A (25 ng/ml). After 24h images of the sprouts were obtained using phase-contrast microscopy. Using ImageJ \cite{Schneider2012} with the NeuronJ plugin \cite{Meijering2004} the number of sprouts and the length of the sprouts in the image were counted.  To compare the \emph{in silico} simulations with the \emph{in vitro} experiments,  \emph{in silico} morphologies at 750 MCS were analyzed following the same method. To prevent biases in this manual analysis due to prior knowledge, black and white images in which tip and stalk cells were indistinguishable (see \nameref{S6_Fig}) were counted by a technician. 

To study sprouting in absence of tip cells, CD34 negative HMEC-1s \cite{Siemerink2012} were extracted using Fluorescence-activated cell sorting (FACS). For this the cells were washed in PBS containing 0.1\% bovine serum albumin. Cells were incubated with anti-CD34-phycoerythrin (anti-CD34-PE; clone QBend-10) and analyzed by flow cytometry on a FACSCalibur (Becton Dickinson, Franklin Lakes, NJ, USA) with FlowJo 6.4.7 software (Tree Star, San Carlos, CA, USA).  

To inhibit Apelin signaling HMEC-1s were transfected with a silencing RNA (siRNA) against Apelin (siAPLN) or against the Apelin receptor APJ (siAPJ), and a non-translating siRNA (siNT) was used as a control. For each siRNA the HMEC-1s were transfected with 25 nM siRNA (Dharmacon, Lafayette, CO, USA) final concentration and 2.5 nM Dharmafect 1 (Dharmacon) for 6 hours using the reversed transfection method \cite{sirna}. Transfection efficiency was evaluated with qPCR and a knockdown of RNA expression above 70\% was considered as an effective transfection.

For both the unsorted HMEC-1s and the CD34 negative HMEC-1s the experiments were repeated several times, resulting in 4 biological replicates for the unsorted HMEC-1s and 5 biological replicates for the CD34 negative HMEC-1s. To combine the results of the biological replicates, the number of sprouts $n^i_R$ of spheroid $i$ in replicate $R$ were normalized: $N^i_R = \frac{n^i_R}{\bar{n}^\text{siNT}_R}$, with $\bar{n}^\text{siNT}_R$ the average number of sprouts formed with the non-translating siRNA treatment in biological replicate $R$. Next, we computed the average number of sprouts per replicate: $\bar{N}_R = \frac{\sum^{m_R} N^i_R }{m_R}$, with $m_R$ the number of spheroids in replicate $R$. This resulted in four data points for the unsorted HMEC-1s and five data points for the CD34 negative HMEC-1s. Then, significance of each treatment was analyzed in a two-step procedure. First, groups in which the means differ significantly were identified with analysis of variance (ANOVA). Second, to identify which means in a group differ, we used Tukey's range test \cite{Tukey1949,Abdi2010} to compare the results of the treatments in plain collagen with the siNT treatment in plain collagen and the treatments in VEGF-enriched collagen with the siNT treatment in VEGF-enriched collagen. All experimental measurements are included in S1 Dataset together with the python script used to perform the statistical analysis. An archive containing the photographs of the HMEC spheroids used for the image analysis is included as~\nameref{S2_Dataset}.

\subsection*{Estimation of endothelial cell cross-sectional area}
The spheroid assay was performed as described above. Gels were fixed with 4\% paraformaldehyde for 15 min at room temperature and blocked with blocking buffer containing 1\% FBS, 3\% triton x-100 (sigma), 0.5\% tween-20 (sigma), 0.15\% natriumazide for 2 hours. Cells where incubated with antibodies directed against F-actin (Phalloidin, Life technologies, Carlsbad, CA, USA). Three-dimensional image stacks were recorded using confocal microscopy. Within those, images containing the largest cross-section were selected visually, and measurements were obtained using the ImageJ polygonal selection tool. The image stacks and measurements are include as~\nameref{S3_Dataset}.

\section*{Supporting Information}


\subsection*{S1 Video}
\label{S1_Video}
{\bf  Cells aggregate instead of forming a network with 20\% predefined tip cells and $J(\text{tip,ECM})=5$.}  

\subsection*{S2 Video}
\label{S2_Video}
{\bf Close up of tip cells on the side of a branch that cause network expansion.} For this simulation the 20\% of the cells were predefined as tip cells with $\chi(\text{tip})=400$. 

\subsection*{S3 Video}
\label{S3_Video}
{\bf  Selected tip cells do not pull apart the network in a simulation with $\Theta_\text{NICD}=0.1$ and $\chi(\text{tip})=200$.}

\subsection*{S4 Video}
\label{S4_Video}
{\bf  Sprouting is strongly inhibited for $\Theta_\text{NICD}=.2$ and 90\% inhibition of Apelin secretion ($\alpha(\text{tip})=10^{-3}\text{s}^{-1}$ and $\alpha(\text{stalk})=10^{-4}\text{s}^{-1}$).}

\subsection*{S5 Video}
\label{S5_Video}
{\bf  When the model is adapted for Apelin, `predefined' tip cells get surrounded by stalk cells.} For this simulation 10\% of the cells were predefined as tip cells with $\chi(\text{tip})=400$ and $\alpha(\text{tip})=0.01$.

\subsection*{S1 Dataset} 
\label{S1_Dataset}
{\bf Archive containing the results of morphological analysis of the \emph{in vitro} endothelial sprouting assays.} The archive contains one text file for each treatment for each replicate and the python script used to perform the statistical analysis.

\subsection*{S2 Dataset}
\label{S2_Dataset}
{\bf Archive containing the photographs of the HMEC-1 spheroids.} Images in TIFF format, as used for the image analysis. The archive also includes the output files of the NeuronJ~\cite{Meijering2004} plugin to ImageJ~\cite{Schneider2012} (file extension ``.NDF'') as well as XML-files containing microscopy settings. The dataset (1.6 GB ZIPped archive) is available via Data Archiving and Networked Services (DANS) at \url{http://dx.doi.org/10.17026/dans-x4d-b642}.

\subsection*{S3 Dataset}
\label{S3_Dataset}
{\bf Archive containing fluorescently stained photographs of individual endothelial cells and whole HMEC-1 spheroids, as used for area estimation.} Images in TIFF format. The measurements (performed using ImageJ) are given in file ``cellareas.xlsx''. 

\subsection*{S1 File}
\label{S1_File}
{\bf Simulation script and code needed to run the simulations in the CPM modeling framework \emph{CompuCell3D} \cite{Swat2012}.} The simulation script (angiogenesis.xml) can be used when the two CC3D steppables, RandomBlobInitializer and TCS, are compiled and installed. RandomBlobInitializer is needed to initialize a simulation with a circular blob and this steppable may be replaced with CC3D's BlobInitalizer. TCS is the steppable that runs the Dll4-Notch genetic network and should be omitted to run simulations with predefined tip cells.

\subsection*{S1 Fig}
\label{S1_Fig}
\includegraphics{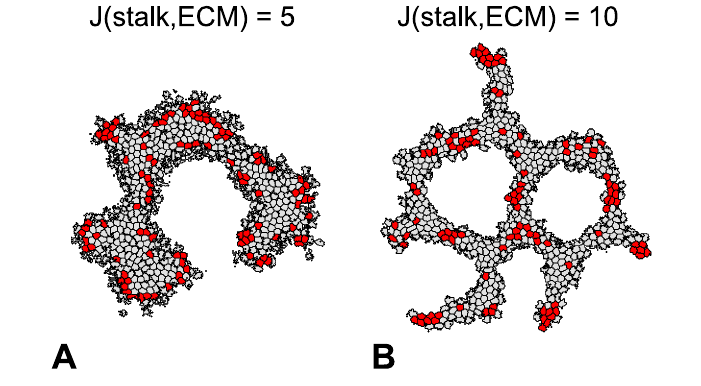}\\
{\bf Effects of increasing ECM adhesion for stalk cells.} \textbf{A} stalk cells that adhere more strongly to the ECM than tip cells will engulf tip cells. \textbf{B} stalk cells that adhere slightly more to the ECM than tip cells do engulf tip cells, because chemotaxis has the same effect on tip and stalk cells. \textbf{A-B} are the results of a simulation of 10 000 MCS with 20\% tip cells.

\subsection*{S2 Fig}
\label{S2_Fig}
\includegraphics[width=\textwidth]{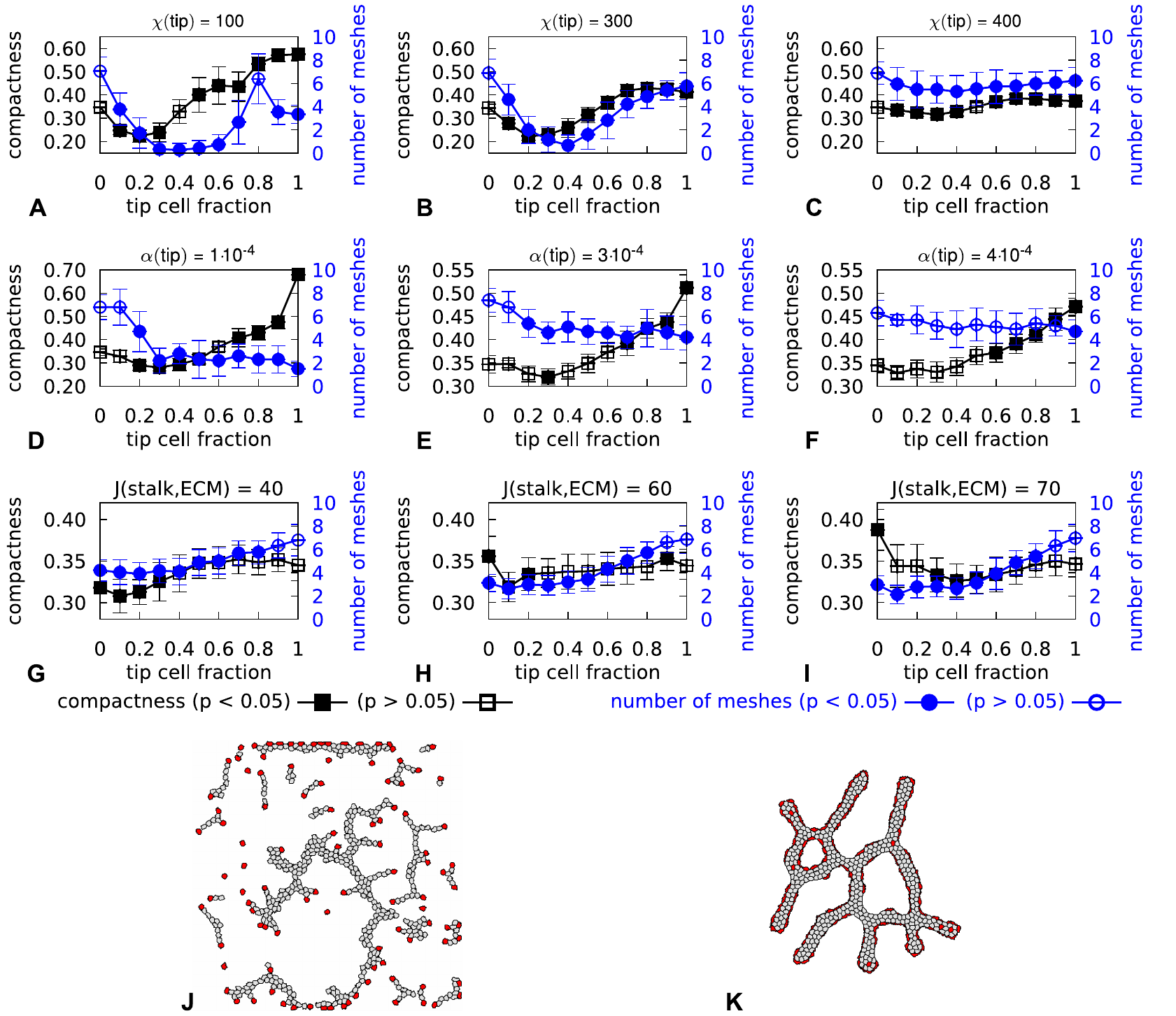}\\
{\bf Effects of varying tip cell chemotaxis.} (\textbf{A}-\textbf{C}), tip cell chemoattractant secretion rate (\textbf{D}-\textbf{F}) and stalk-ECM adhesion (\textbf{G}-\textbf{I}). The morphometrics were obtained after 10 000 MCS and are the average of 50 simulations (error bars represent standard deviation). p-values were obtained with a Welch's t-test for the null hypothesis that the mean of the sample is identical to that of a reference where all cells have the default properties. \textbf{J} the network disintegrates with $\chi(\text{tip}) = 100$) and 20\% tip cells. \textbf{K} tip cells over the network for $J(\text{stalk,ECM})=70$ and 20\% tip cells.

\subsection*{S3 Fig}
\label{S3_Fig}
\begin{center}
\includegraphics[width=.75\textwidth]{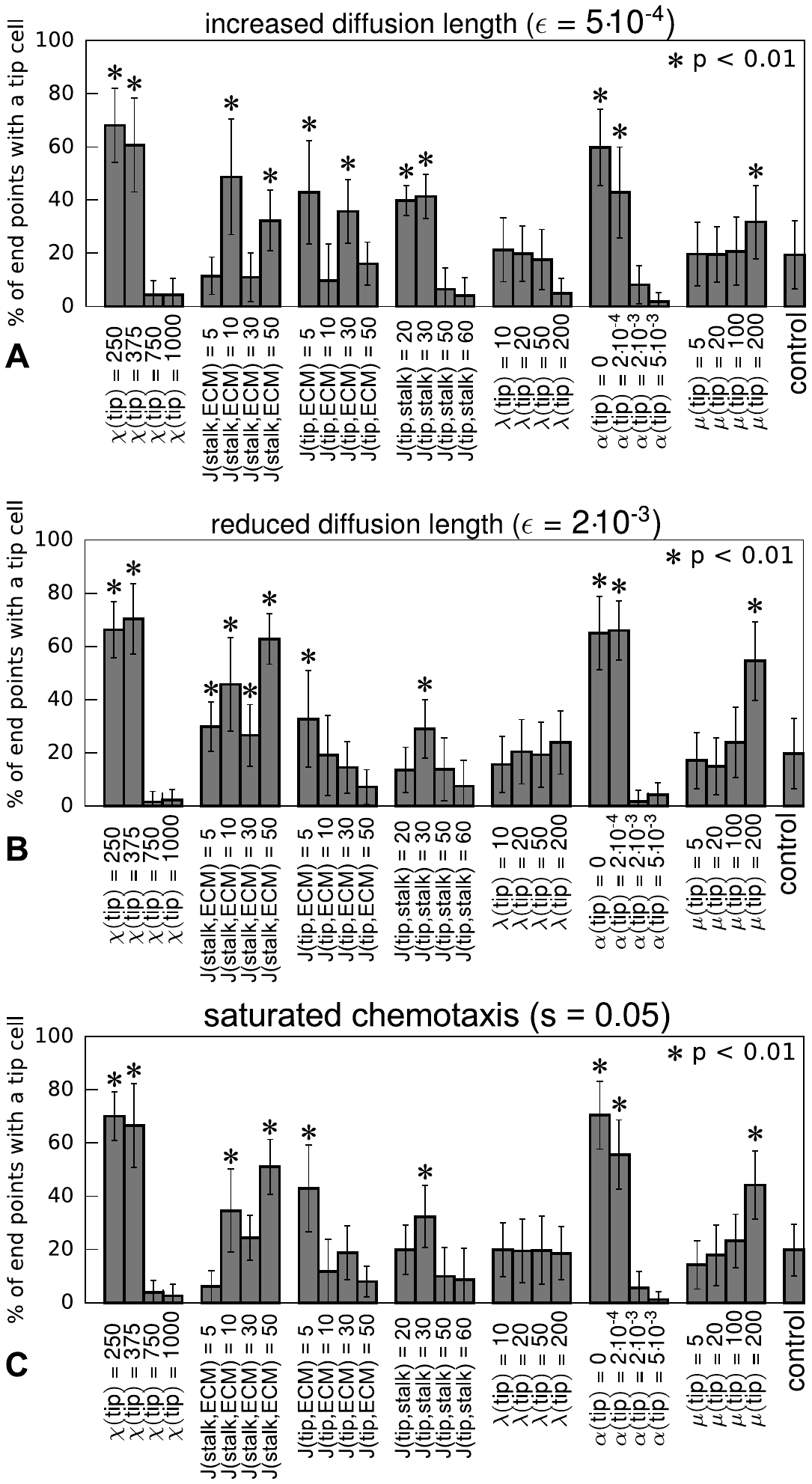}\\
\end{center}
{\bf Differences in cell properties can enable cells of one type to occupy sprout tips for three alternative parameter sets.} For \textbf{A} the decay rate was reduced, for \textbf{B} the decay rate was increased and for \textbf{C} receptor saturation was included in the model. The percentage of sprout tips occupied by at least one tip cell was calculated at 10 000 MCS. Error bars show the standard deviation over 50 simulations. In each simulation 20\% of the cells were predefined as tip cells. For each simulation one tip cell parameter was changed, except for the control experiment where the nominal parameters were used for both tip and stalk cells. p-values were obtained with a one sided Welch's t-test for the null hypothesis that the number of tip cells at the sprout tips is not larger than in the control simulation.

\subsection*{S4 Fig}
\label{S4_Fig}
\includegraphics[width=\textwidth]{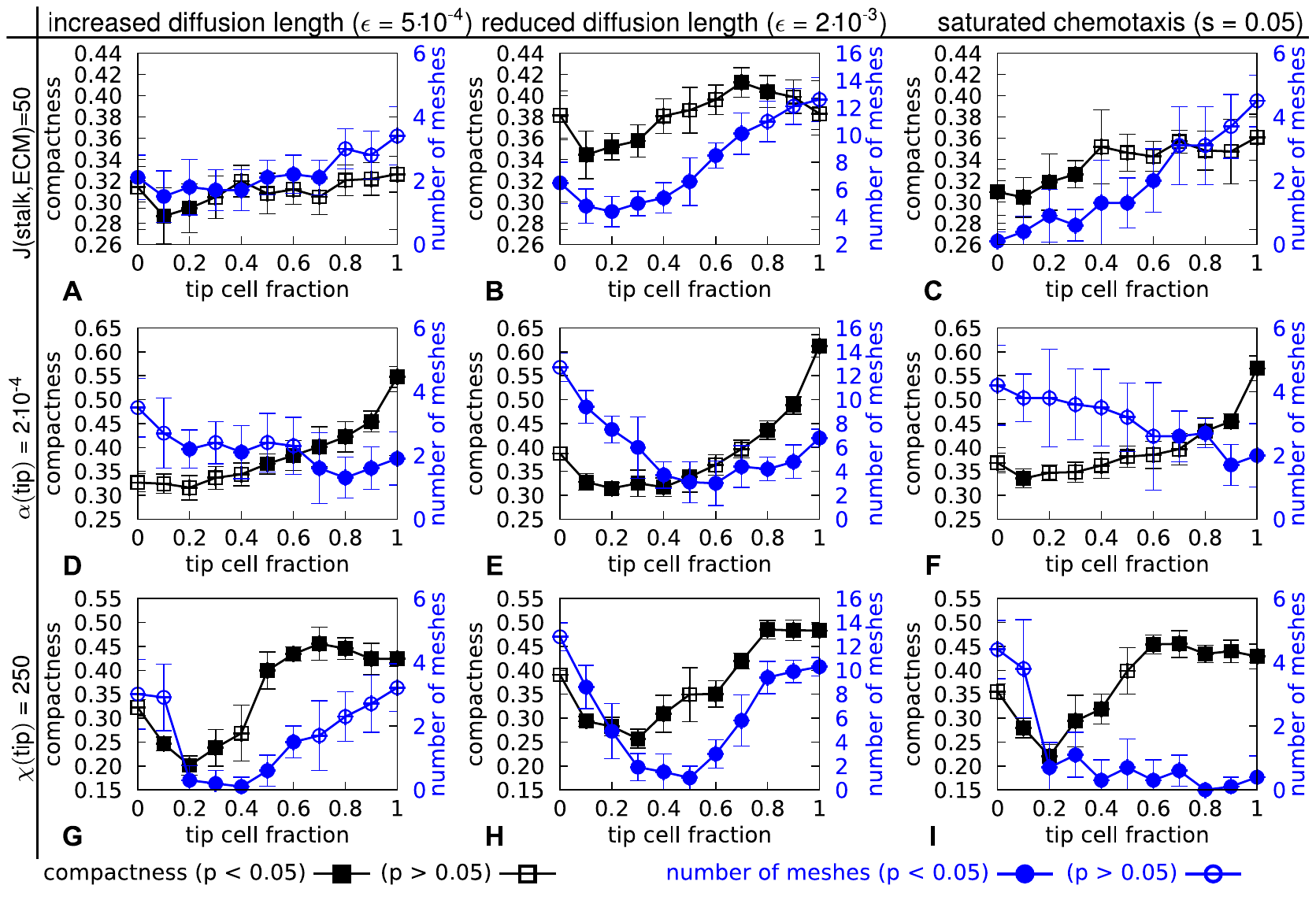}\\
{\bf Effects of tip cells with $J(\text{stalk,ECM}$.} (\textbf{A}-\textbf{C}), $\alpha(\text{tip})$ (\textbf{D}-\textbf{F}) or $\chi(\text{tip})$ (\textbf{G}-\textbf{I}) on the network morphology for the three alternative parameter sets. The morphometrics were obtained after 10 000 MCS and are the average of 10 simulations (error bars represent standard deviation).  p-values were obtained with a Welch's t-test for the null hypothesis that the mean of the sample is identical to that of a reference sample in which all cells have the default properties.

\subsection*{S5 Fig}
\label{S5_Fig}
\includegraphics[width=\textwidth]{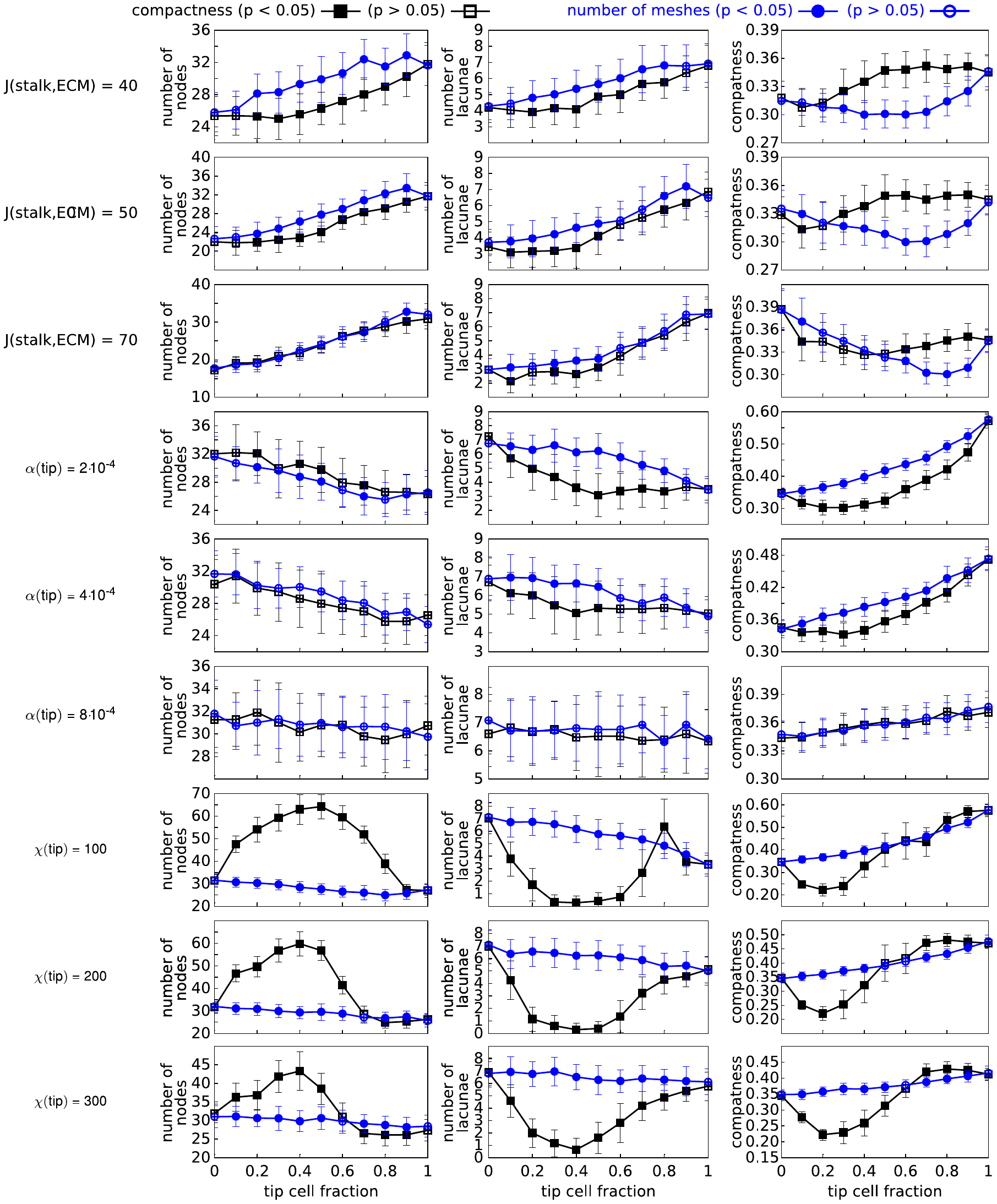}\\
{\bf Comparison of networks formed with mixed cells and cells with average properties for additional value of $J(\text{stalk,ECM})$, $\alpha(\text{tip})$ and $\chi(\text{tip})$.} The morphometrics were calculated for 50 simulations at 10 000 MCS (error bars represent the standard deviation). p-values were obtained with a Welch's t-test for the null hypothesis that the mean of mixed model and the control model are identical.

\subsection*{S6 Fig}
\label{S6_Fig}
\begin{center}
\includegraphics{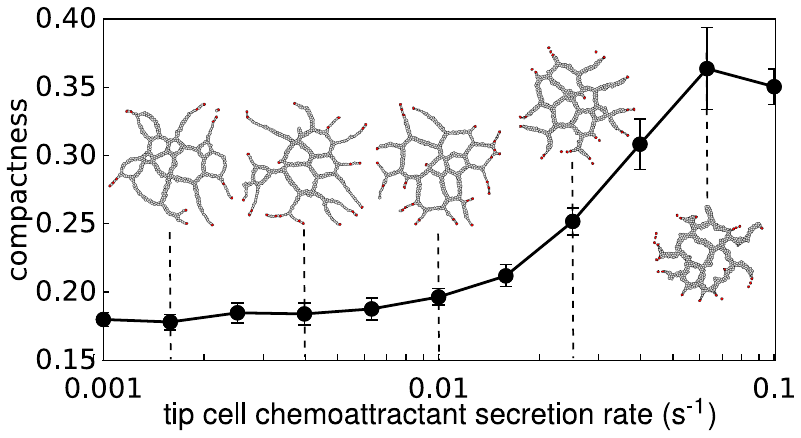}\\
\end{center}
{\bf Effects of increasing tip cell Apelin secretion rate for tip cells that do not respond to Apelin.} Compactness of the final network (10 000 MCS) with the morphologies for for tip cell Apelin secretion rates of $\alpha(\text{tip})=1.6\cdot 10^{-3}$, $\alpha(\text{tip})=4.0\cdot 10^{-3}$, $\alpha(\text{tip})=1\cdot 10^{-2}$,$\alpha(\text{tip})=2.5\cdot 10^{-2}$, and $\alpha(\text{tip})=6.3\cdot 10^{-2}$ as insets. To enable network formation without tip cell chemotaxis $J(\text{tip,tip})$, $J(\text{stalk,stalk})$ and $J(\text{tip,stalk})$ were reduced to 25. Data points show average values for $n=50$ simulations with error bars giving the standard deviation.

\subsection*{S7 Fig}
\label{S7_Fig}
\includegraphics[width=\textwidth]{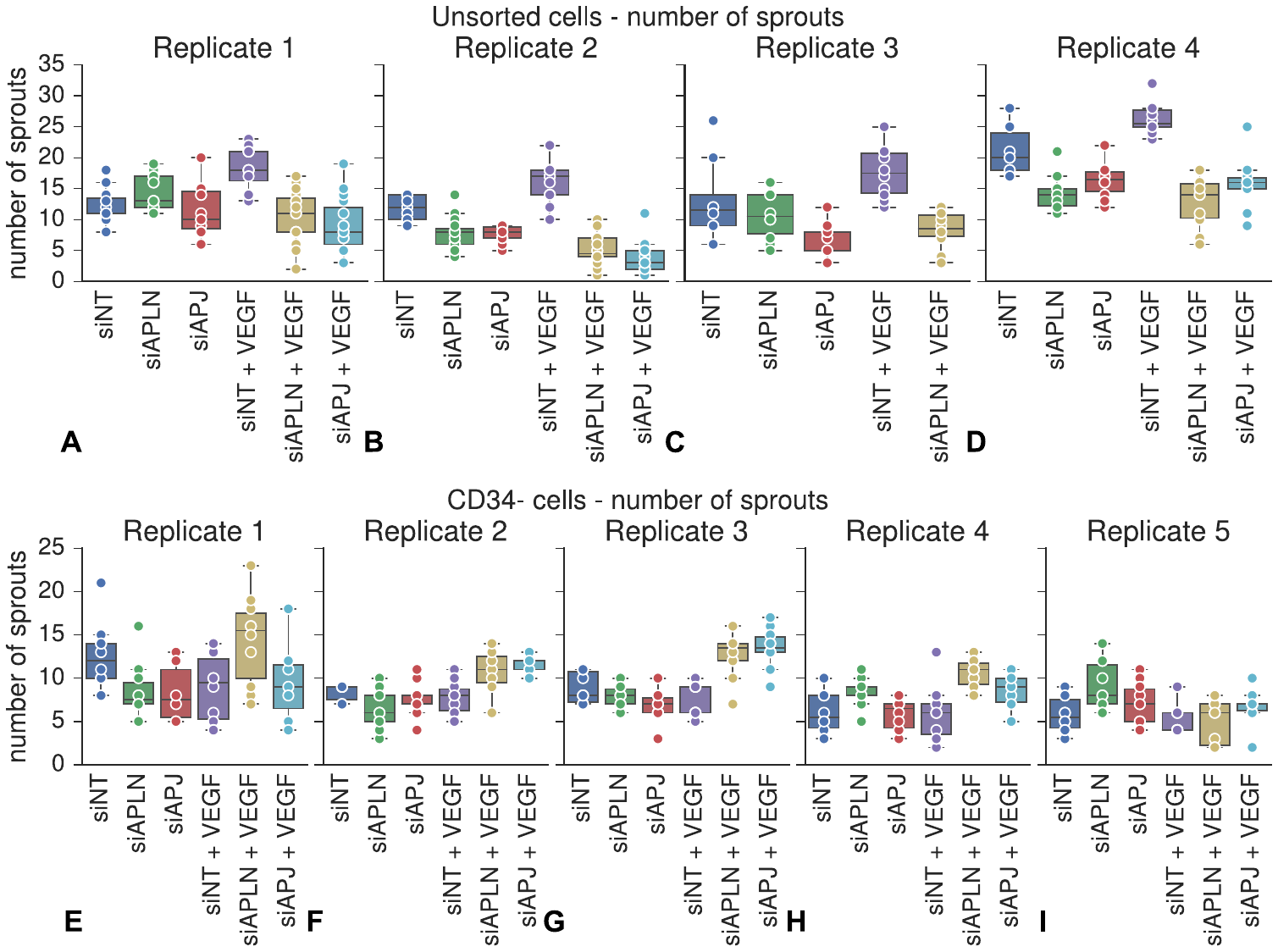}\\
{\bf Effect of siAPJ and siAPLN on sprout lengths for all experiments.} The boxes show the first to third quartile of the data. The whiskers show Q1-1.5$\cdot$IQR to Q3+1.5$\cdot$IQR, with IQR = Q3-Q1 = interquartile range, or the most extreme observations if those fall within the range of the whisker. Superimposed on the box plots are the data points. Note that the experiments are done with different collagen gels: Purecol collagen (\textbf{A},\textbf{E}), Nutacon collagen (\textbf{B}, \textbf{F}, \textbf{G}), and Cultrex rat collagen (\textbf{C}, \textbf{D}, \textbf{H}, \textbf{I}).

\subsection*{S8 Fig}
\label{S8_Fig}
\includegraphics[width=\textwidth]{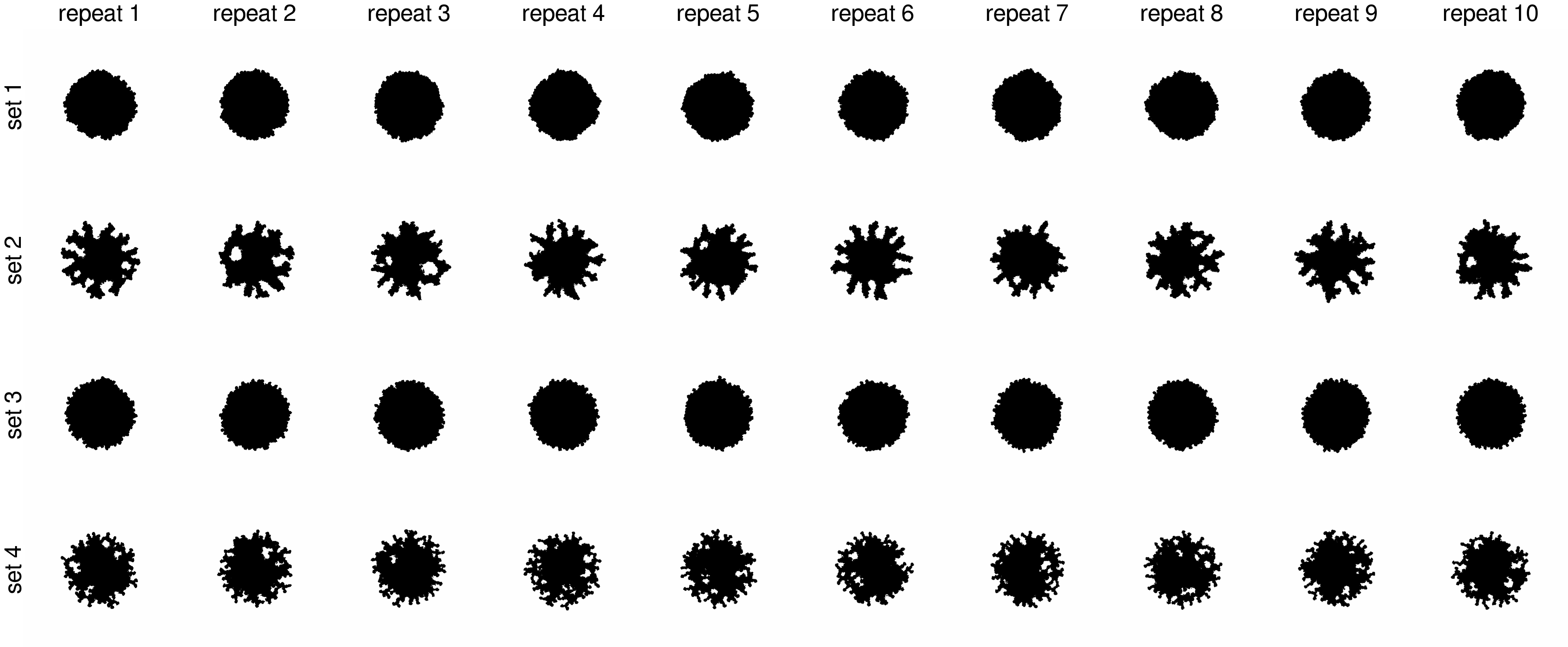}\\
{\bf Black and white images of the morphologies produced in the \emph{in silico} Apelin silencing experiments as provided to the technician.}


\section*{Acknowledgments}
The authors thank Indiana University Bloomington and the Biocomplexity Institute for providing the CC3D modeling environment (\url{www.compucell3d.org}) \cite{Swat2012}. This work was carried out on the Dutch national e-infrastructure with the support of SURF Cooperative (\url{www.surfsara.nl}).


\end{document}